\documentclass[preprint,12pt]{elsarticle}

\usepackage{amssymb}
\usepackage{amsmath}
\usepackage{graphicx} 
\usepackage[utf8]{inputenc}
\usepackage{multirow}
\DeclareUnicodeCharacter{2212}{-}
\usepackage{tikz}
\usepackage{tabularx}
\usepackage[table]{xcolor}
\usepackage{subcaption} 
\usepackage[most]{tcolorbox}
\usepackage{booktabs}
\usepackage[ruled,vlined]{algorithm2e}
\usepackage{enumitem}
\usepackage{fbox}
\usepackage{url}
\usepackage[section]{placeins}
\usepackage{tikz}
\usepackage{adjustbox}
\usetikzlibrary{shapes.geometric}

\newcommand{\llamamarker}{\tikz[baseline=-0.5ex]\node[circle, fill=blue, inner sep=1.2pt]{};}
\newcommand{\gemmamarker}{\tikz[baseline=-0.5ex]\node[rectangle, fill=orange, inner sep=1.2pt]{};}
\newcommand{\qwenmarker}{\tikz[baseline=-0.5ex]\node[regular polygon, regular polygon sides=3, fill=green, inner sep=1.2pt]{};}
\newcommand{\mistralmarker}{\tikz[baseline=-0.5ex]\node[regular polygon, regular polygon sides=3, fill=red, inner sep=1.2pt, rotate=180]{};}
\newcommand{\redcircle}{\tikz[baseline=-0.5ex]\node[circle, fill=red, inner sep=1.2pt]{};}
\newcommand{\greensquare}{\tikz[baseline=-0.5ex]\node[rectangle, fill=green, inner sep=1.2pt]{};}

\journal{Nuclear Physics B}

\begin{document}

\begin{frontmatter}



\title{
Fairness-Aware Retrieval Optimization for Retrieval-Augmented Generation
} 

\author[label1]{Yingqi Zhao}
\ead{yingqi.zhao@tuni.fi}

\author[label2]{Vasilis Efthymiou}
\ead{vefthym@hua.gr}

\author[label1]{Jyrki Nummenmaa}
\ead{jyrki.nummenmaa@tuni.fi}

\author[label1]{Kostas Stefanidis}
\ead{konstantinos.stefanidis@tuni.fi}

\affiliation[label1]{organization={Data Science Research Centre, Tampere University},
            city={Tampere},
            country={Finland}}

\affiliation[label2]{organization={Harokopio University of Athens},
            city={Athens},
            country={Greece}}

\begin{abstract} 
Retrieval-Augmented Generation (RAG) grounds large language models (LLMs) in external knowledge, but the retrieved documents can systematically skew generated answers toward one group. This effect is difficult to analyze and control in top-$k$ settings, where several documents at different context positions jointly shape the output. We present a three-stage framework for fairness-aware retrieval in top-$k$ RAG. First, a probabilistic reranker sets the group composition of the retrieved context at each position, enabling controlled perturbation of retrieval bias without modifying the retriever. Second, we show that output bias is well approximated by a position-weighted linear function of position-wise retrieval bias, revealing model-specific positional sensitivities. Third, we use this model to formulate retrieval as an optimization problem that trades off relevance against predicted output bias, and propose FARO, a scalable solution to that problem which decomposes the globally coupled quadratic objective into independent per-query assignment problems through a Fenchel--Legendre dual reformulation. Experiments with four open LLMs, and political and gender bias as use cases show that the linear model predicts output bias on unseen configurations 
in 17 of 24 settings. FARO substantially reduces output bias in most settings, while,
compared with a linear-programming baseline, FARO achieves similar relevance, remains feasible where the baseline does not, and scales better to larger query sets and candidate pools.
\end{abstract}




\begin{keyword}
Retrieval-Augmented Generation (RAG) \sep Fairness-Aware Retrieval \sep Bias Propagation \sep Fair Ranking.

\end{keyword}

\end{frontmatter}



\section{Introduction} 
Recent advances in Large Language Models (LLMs) have significantly improved the ability of artificial intelligence systems to generate coherent and contextually relevant text~\cite{vaswani_attention_2017,kaplan_scaling_2020}. Despite their success, LLMs often suffer from hallucinations and factual inconsistencies, limiting their reliability in real-world applications~\cite{Ji_Hallucination_2023}. Retrieval-Augmented Generation (RAG)~\cite{Lewis_RAG_2020} has emerged as an effective paradigm to address these limitations by incorporating external knowledge into the generation process. By retrieving relevant documents and feeding them to the language model as a part of the input, RAG enhances factual grounding and improves answer quality. 

However, while RAG mitigates hallucinations, it also introduces new challenges related to fairness and bias. In particular, the selection and ordering of retrieved documents can significantly influence the generated output, potentially amplifying biases present in the underlying knowledge sources or retrieval mechanisms. As a result, even when the language model itself does not exhibit significant bias, biased retrieval can lead to systematically skewed responses~\cite{hu_no_2024,wu_does_2025}. This issue is especially pronounced in top-$k$ RAG settings, where multiple ($k$) documents jointly contribute to the final output, creating complex interactions that are difficult to analyze and control.

Existing research has explored bias in LLMs \cite{DBLP:journals/corr/abs-2409-16430} and fairness in ranking systems \cite{DBLP:journals/vldb/PitouraSK22}, but these lines of work remain largely disconnected. Studies on LLM bias primarily focus on measuring or mitigating bias through prompt engineering, fine-tuning, or dataset curation, without explicitly considering the potential of RAG~\cite{gallegosBiasFairnessLarge2024c}. Conversely, fairness-aware ranking methods aim to ensure equitable exposure of items or providers, but do not account for how ranking decisions affect downstream text generation~\cite{singhFairnessExposureRankings2018}. Consequently, there is a lack of principled approaches for controlling generation bias through retrieval decisions in RAG systems. 

\begin{figure}[t]
    \centering
    \includegraphics[width=1.0\linewidth]{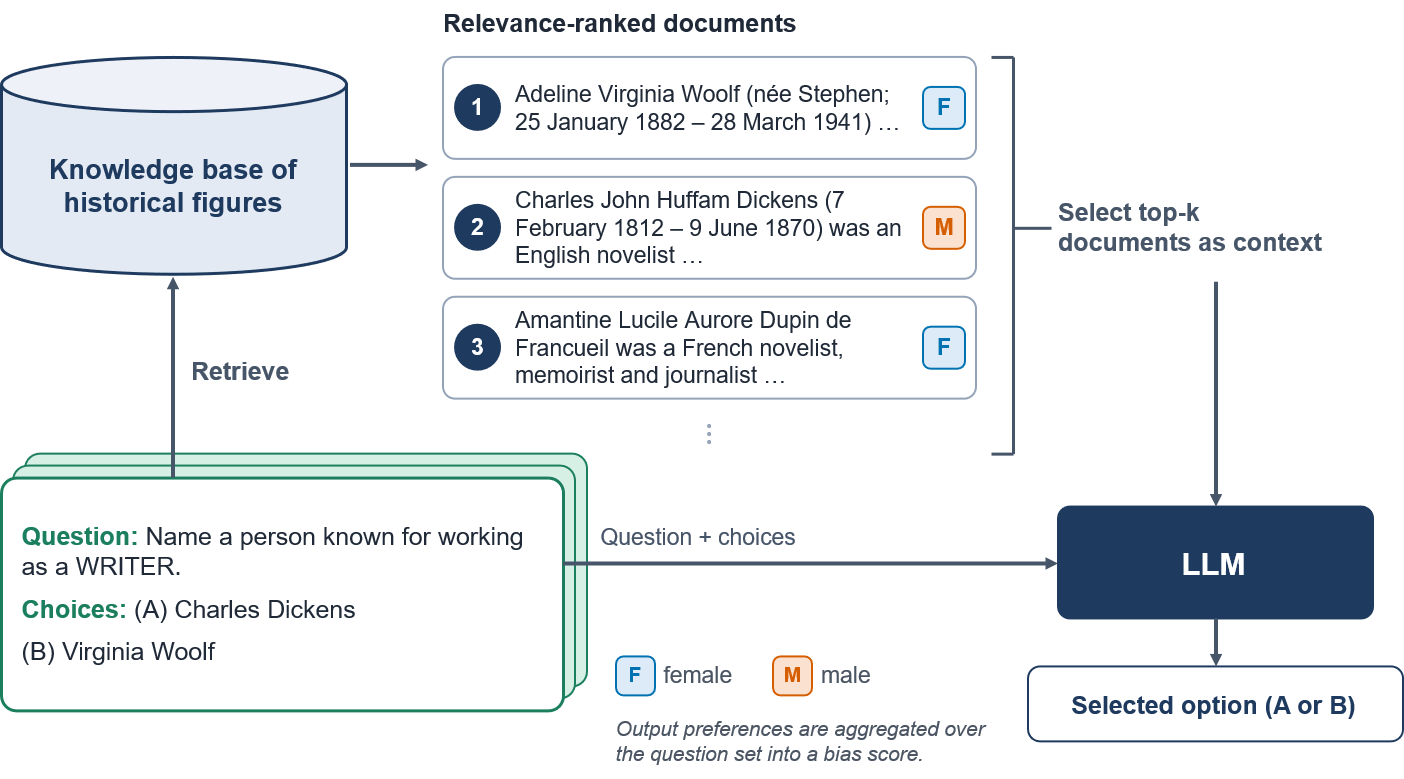} 
    
    \caption{An illustration of top-$k$ RAG for gender bias evaluation. The question offers a male and a female candidate, both of whom satisfy the stated occupation. The LLM selects an answer using $k$ ranked documents retrieved from a knowledge base of historical figures. The gender composition and ordering of these documents may influence answer selection. Gender bias is evaluated by aggregating output preferences across the question set.}
    \label{fig:example}
\end{figure}

A key challenge lies in understanding how bias propagates from retrieved documents to generated outputs. While prior work has identified a linear relationship between embedding bias and output bias in top-1 settings~\cite{kim_mitigating_2025,zhaoReFaRAGRerankingBias2026}, this assumption does not directly extend to top-$k$ retrieval, where multiple documents with different positions jointly influence the generation process, as shown in the example in Figure~\ref{fig:example}. Modeling this position-dependent interaction is essential for designing effective fairness-aware retrieval strategies. 

In this work, we address this challenge by introducing a unified framework for fairness-aware retrieval in top-$k$ RAG systems. Our approach is based on the insight that generation bias can be controlled by manipulating the bias of retrieved documents and understanding how this bias propagates through the model. To this end, we propose a three-stage framework. First, we extend our reranker-based mechanism that enables controlled bias injection by adjusting the group distribution of retrieved documents at different ranking. Second, we develop a position-aware bias propagation model that captures how bias at different retrieval positions influences the final output. Third, we formulate fairness-aware retrieval as an optimization problem and propose an efficient solution that balances relevance and fairness.

Building on this formulation, we first define the Fairness-Aware Retrieval Optimization problem, which balances retrieval relevance and predicted generation bias. We then propose a scalable algorithm, termed FARO, to solve this problem. FARO transforms the globally coupled optimization problem into a set of independent subproblems, enabling efficient computation while allowing trade-offs between relevance and fairness. Using the learned bias propagation model, FARO links retrieval decisions to predicted generation bias, providing a principled mechanism for bias control in RAG systems. 

We evaluate our approach on multiple datasets and models, considering both political and gender bias settings. The experimental results demonstrate that our method effectively reduces generation bias while maintaining competitive relevance performance. Furthermore, the proposed optimization framework offers flexibility in navigating the relevance–fairness trade-off.

Overall, we propose an end-to-end three-stage framework for fairness-aware retrieval optimization in RAG systems. Our approach integrates controlled bias injection, position-aware modeling of bias propagation, and fairness-aware optimization of retrieval decisions. The framework provides a principled and scalable solution for balancing relevance and fairness in top-$k$ RAG settings. The main contributions of this work are: 
\begin{itemize} 
\item We extend reranking-based bias control to individual ranking positions. We show that output bias is well approximated by a position-weighted linear function of position-wise retrieval bias. The learned weights reveal model-specific positional sensitivities: the influence of any single position decreases as $k$ grows, and some positions carry negative weight. 
\item We introduce fairness-aware retrieval optimization and the FARO algorithm. We formulate retrieval as relevance maximization under a dataset-level constraint on predicted output bias. To overcome the global coupling this constraint introduces, FARO applies a Fenchel-Legendre dual reformulation of a quadratic fairness penalty. This reduces the problem to independent per-query assignment problems that are solved exactly with the Hungarian algorithm, and it yields a family of candidate solutions along the relevance–fairness frontier. 
\item We conduct extensive experiments. We experiment with four open LLMs, two bias types (political and gender), and different $k$ values. The experiments validate the propagation model through out-of-fold prediction and cross-dataset transfer. They also show that FARO substantially reduces output bias in most settings, achieves relevance similar to a linear programming baseline, remains feasible where the baseline does not, and scales better with the number of queries and candidate documents. 
\end{itemize}  

Compared with our earlier work~\cite{zhaoReFaRAGRerankingBias2026}, which studied bias propagation for top-1 retrieval and political bias only, this paper adds the position-aware top-$k$ propagation model, the optimization formulation and FARO, the gender-bias experiments across four LLMs (using the dataset of \cite{zhao2026dataset}), a new political-bias question set with cross-dataset validation, and the scalability analysis.

The remainder of the paper is structured around the proposed fairness-aware retrieval pipeline. Section \ref{sec:relw} reviews related work on bias in RAG systems and fairness-aware ranking. Section \ref{sec:overview} introduces the overall three-stage framework and formalizes the problem setting. Section \ref{sec:fairtopkrag} develops a position-aware model of bias propagation, capturing how retrieved documents jointly influence generated outputs. Section \ref{Optimization-Based Fair Retrieval in RAG} formulates fairness-aware retrieval as an optimization problem and presents the FARO for scalable solution. Section \ref{section:expsetup} presents the experimental setup, and Section \ref{sec:expresults} reports experimental results, including validation of the bias propagation model and evaluation of the proposed optimization approach. Finally, Section \ref{sec:conclusion} concludes the paper with a summary of our contributions.

\section{Related Work} 
\label{sec:relw}
\subsection{Bias Introduction and Mitigation in RAG}
Although Retrieval-Augmented Generation (RAG) has been widely adopted to improve the performance of Large Language Models (LLMs) and reduce hallucinations, recent studies have shown that it may also introduce or amplify bias in generated outputs. 
Hu et al.~\cite{hu_no_2024} propose a three-stage fairness-aware framework by controlling the proportion of biased content in the knowledge base. Their findings show that even a small amount of biased information can significantly influence model outputs, and that RAG may amplify such biases. Moreover, even when retrieved knowledge appears unbiased, it can still degrade the alignment behavior of LLMs. 

Wu et al.~\cite{wu_does_2025} conduct extensive ablation studies to analyze how different RAG components, such as the retriever, generator, refiner, and judger, affect bias in generated outputs. Their results indicate that both the retriever and the generator are primary contributors to bias, highlighting the importance of interactions across components. Similarly, Zhang et al.~\cite{zhangEvaluatingEffectRetrieval2025} demonstrate that retrieval strategies have a significant impact on bias, and show that this effect persists across multiple languages, including Chinese and Japanese.

A key step toward understanding bias in RAG is the identification of bias propagation mechanisms. Kim et al.~\cite{kim_mitigating_2025} are the first to reveal a linear relationship between embedding bias and output bias in a top-1 RAG setting. By fine-tuning the retriever, they control the bias distribution of retrieved documents and, through this relationship, influence the downstream generation behavior. 
However, this approach presents important limitations. On the one hand, reliance on black-box embedding models introduces additional uncertainty and makes it difficult to achieve precise control over retrieval outcomes, even in the top-1 setting, while also limiting extension to top-$k$ retrieval scenarios. On the other hand, fine-tuning the embedding model introduces substantial cost, requiring dataset construction and retriever training, as well as repeated experiments to estimate the bias relationship. 


To address these limitations, our previous work~\cite{zhaoReFaRAGRerankingBias2026} proposes a reranker-based approach that directly controls embedding bias through probabilistic selection. This method reduces deployment cost, simplifies implementation, and enables more precise manipulation of bias in retrieved documents. 
Building upon this line of work, we extend the analysis from top-1 to top-$k$ RAG settings. This extension is crucial for real-world applications, where multiple documents jointly influence the generated output. While prior work (e.g.,~\cite{liu_lost_2024}) has studied attention patterns in long-context LLMs, relatively little research has examined how bias propagates across multiple retrieved documents. In contrast, our work explicitly models position-dependent bias propagation and leverages it for bias mitigation.

It is worth noting that there are other works on fairness-aware ranking in RAG, such as~\cite{kim_towardsfarirRAG_2025}, focusing on the fair presentation of retrieved relevant items, thereby enabling the generator to reference its sources in a balanced manner;~\cite{DehghanWho2026} instead examines whether question associated with certain groups within a specific fairness category systematically achieve higher accuracy, and studies the roles of exposure, utility, and attribution bias. These approaches define and evaluate fairness in RAG along a different dimension from ours, and we therefore do not discuss them here.

\subsection{Fairness-Aware Ranking and Optimization} 
The optimization framework proposed in this work is related to prior studies on fairness-aware ranking, particularly in recommender systems and information retrieval. 
FA*IR~\cite{zehlike_fair_2017} introduces one of the earliest approaches to fair top-$k$ ranking, enforcing statistical fairness constraints over binary groups. Singh et al.~\cite{singhFairnessExposureRankings2018} further highlight that ranking positions induce unequal exposure due to position bias, and propose aligning exposure with relevance through probabilistic ranking formulations. DELTR~\cite{zehlikeReducingDisparateExposure2020} incorporates fairness as a regularization term within learning-to-rank objectives, enabling fairness-aware training. Additional works, such as~\cite{beutelFairnessRecommendationRanking2019} and~\cite{singhPolicyLearningFairness2019}, extend this idea by integrating fairness constraints into ranking objectives using pairwise comparisons or reinforcement learning.

Despite their relevance, these approaches are not directly applicable to RAG systems due to fundamental differences in objectives. In recommenders, fairness is typically defined in terms of exposure allocation to items or providers, whereas in RAG the primary concern is the fairness of the generated outputs. Retrieved documents do not directly correspond to entities requiring fair treatment; instead, they influence the generation process of the LLM.

Nevertheless, these works provide important conceptual foundations. In particular, the notion of position-dependent exposure aligns with our observation that documents at different retrieval positions have varying influence on the final output. This perspective allows us to reinterpret retrieval in RAG as a ranking problem under fairness constraints, where the objective is to balance relevance with the bias induced in the generated responses. 
Building on this insight, we leverage the empirically identified linear bias propagation model to connect retrieval decisions with generation bias. This enables the formulation of a fairness-aware retrieval optimization problem tailored to RAG systems, which we address in Section~\ref{Optimization-Based Fair Retrieval in RAG}.

\section{Overview of the Three-Stage Framework for Fairness-Aware Retrieval Optimization in RAG}
\label{sec:overview}
\subsection{Fairness in RAG Systems: Problem Setting}
\label{sec:The starting point for fairness in the RAG system}
The notion of fairness adopted in this work is grounded in a statistical perspective on model outputs. Specifically, we consider a set of questions designed not to favor either candidate answer group, each associated with two candidate answers from groups A and B. We focus on two representative bias types: political bias and gender bias. For political bias, groups A and B correspond to right-oriented and left-oriented viewpoints, respectively; for gender bias, they correspond to female and male representations, respectively. The gender setting defines the scope of our evaluation and is not intended as a comprehensive characterization of gender. In both settings, the questions do not admit a unique correct answer, and both options are equally plausible within the evaluation design. The RAG system is instructed to select one of the two options. Section~\ref{dataset} describes the dataset construction process and provides examples.

Within this controlled evaluation, we operationalize fairness as approximate balance in the aggregate selection frequencies of answers associated with groups A and B. This target follows from the design of the questions and paired answer options; it is not a general requirement that arbitrary questions should produce equal group frequencies. Systematic imbalance therefore indicates directional preference under this evaluation protocol, rather than establishing unfairness in every individual response.

We distinguish directional preference from willingness to answer. A refusal contributes zero to the directional bias score because it favors neither group, but it is not treated as evidence of fairness. In particular, a system that refuses every question has zero measured directional bias without providing any substantive answers. Accordingly, output bias should be interpreted alongside answer coverage and, where available, group preference conditional on a valid selection.

We adopt aggregate directional output bias as the fairness criterion because it provides a measurable link between retrieval decisions and group preferences in generated answers. This enables us to formulate a tractable constraint for controlling retrieval-induced output imbalance while preserving document relevance. The constraint addresses a specific aspect of fairness within the evaluated question set; satisfying it alone does not establish adequate answer coverage, individual fairness, causal fairness, or equitable refusal behavior.

\subsection{Framework Overview} 
We propose a three-stage framework for fairness-aware retrieval optimization in RAG systems. The framework integrates controlled bias injection, bias propagation modeling, and optimization-based retrieval, enabling both the analysis and mitigation of bias in top-$k$ RAG settings.

\begin{figure}[t]
    \centering
    \includegraphics[width=1.0\linewidth]{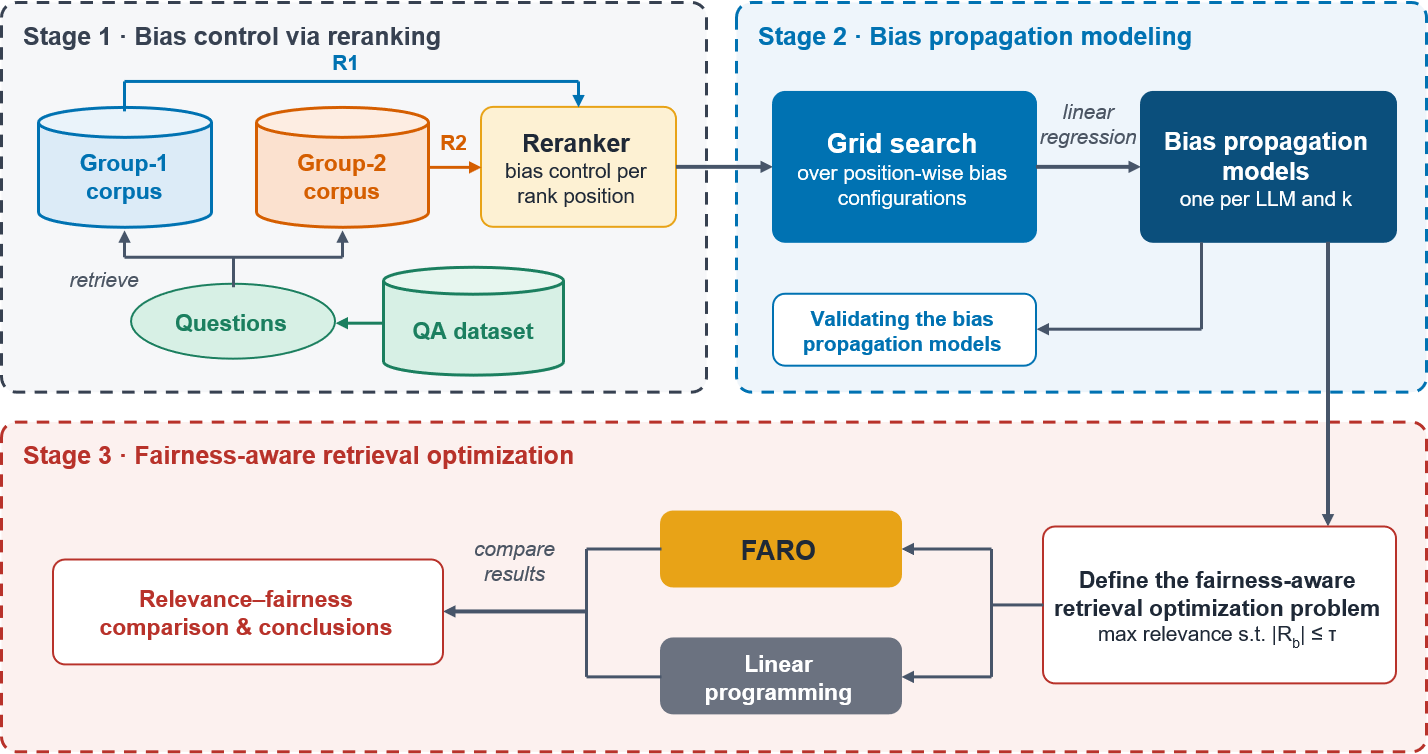} 
    \caption{Overview of the proposed fairness-aware retrieval framework for top-$k$ RAG.
    Stage 1 controls the group composition of retrieved contexts at each rank position through probabilistic reranking.
    Stage 2 explores position-wise bias configurations through grid search and fits a linear bias-propagation model for each LLM and each value of $k$, characterizing how retrieval bias influences generation bias.
    Stage 3 uses the learned models to formulate a retrieval optimization problem that maximizes relevance subject to an aggregate output-bias constraint, $|R_b| \leq \tau$, where $R_b$ denotes the aggregate output bias and $\tau$ is
    the maximum allowed bias magnitude.
    The proposed FARO algorithm is compared with linear programming in terms of relevance, fairness, and computational efficiency.}
    \label{fig:main}
\end{figure}

The key novelty of our approach lies in extending bias control from top-1 retrieval to position-aware top-$k$ retrieval, modeling how bias propagates across multiple retrieved documents, and leveraging this model to guide fairness-aware retrieval optimization. 
An overview of the framework is illustrated in Figure~\ref{fig:main}.

Our framework consists of three stages. First, we introduce a probabilistic reranking controller to control the group composition of retrieved documents at each context position. Second, we apply a grid of probability settings to systematically perturb the retrieved contexts and collect the corresponding RAG outputs. These results enable us to learn and analyze bias propagation models across different LLMs and retrieval settings. Third, we use the learned propagation models to impose fairness constraints and define the fairness-aware retrieval optimization problem. We then propose FARO as a scalable algorithm for solving this problem and compare its performance with a conventional linear programming (LP) baseline. 

Given a set of bias evaluation questions and group-partitioned knowledge bases, the framework produces retrieval strategies that balance relevance and fairness in the final generated outputs.

\subsubsection{Stage 1: Controlled Bias Injection via Reranking} 
The first stage introduces a reranking mechanism to explicitly control the bias of retrieved documents. RAG systems typically construct a knowledge base containing rich and relevant documents, use the input question as a query to retrieve related documents from the knowledge base, and then aggregate their content as contextual information injected into the LLM to improve generation quality. The central idea is to treat the bias of the retrieved context, referred to as embedding bias $E_b$, as a controllable variable that influences the final generation bias $R_b$ of the system.

To enable this control, we partition the knowledge base into group-specific subsets (e.g., liberal vs. conservative, or female vs. male). For each question, we independently retrieve candidate documents from each group-specific subset. A probabilistic reranker is then applied to select and order documents from these subsets when constructing the top-$k$ retrieval list.


Unlike approaches that rely on fine-tuning the embedding model, this design enables direct and precise manipulation of the bias distribution of retrieved documents (for example, the proportion of liberal and conservative viewpoints appearing at specific positions in the retrieval list for political bias) without modifying the underlying retriever. As a result, bias control is decoupled from the retrieval model and can be implemented as a lightweight post-processing step when sufficient group-related documents are available.

Importantly, this mechanism allows us to control not only the overall bias of the retrieved set, but also the bias at each position in the ranked list. We denote the position-wise embedding bias at rank $p$ as $E_b^p$, which serves as the key variable for analyzing bias propagation in subsequent stages. 

\subsubsection{Stage 2: Bias Propagation Modeling} 
The second stage aims to detect how bias in retrieved documents propagates to the final generated outputs. In particular, we seek to quantify the contribution of each position in the retrieval list to the overall system bias. 
To achieve this, we treat the RAG system as a black box and perform controlled perturbations of position-wise embedding bias. Specifically, we generate multiple retrieval configurations by systematically varying the bias values $E_b^p$ across positions in top-$k$ retrieval. For each configuration, we run the RAG system on a set of questions and measure the resulting output bias~$R_b$. 

Using the collected data, we fit a linear regression model that relates position-wise embedding bias to the final system bias. This model captures how bias at different positions contributes to the generated output and provides an interpretable representation of bias propagation. 
The resulting formulation reveals that system-level bias can be approximated as a weighted combination of position-wise embedding biases. The learned coefficients reflect the sensitivity of the language model to bias signals at different positions, and can be interpreted as position-dependent sensitivity weights. 

This modeling step is critical: it provides a tractable and interpretable approximation of how multiple retrieved documents jointly influence the final output, and serves as the foundation for the optimization stage.

\subsubsection{Stage 3: Fairness-Aware Retrieval Optimization}
In the final stage, we leverage the learned bias propagation model to optimize retrieval under fairness constraints. 
The key idea is to treat retrieval as a ranking problem that balances two objectives: maximizing relevance and controlling the bias of the final generated output. Using the linear bias propagation model, we can express the overall system bias as a function of the selected documents and their positions in the ranking. The detailed settings and mathematical formulations are presented in Section~\ref{Optimization-Based Fair Retrieval in RAG}. 

Based on this formulation, we define a fairness constraint that requires the system-level bias to remain within a predefined tolerance. At the same time, we aim to maximize the relevance of the retrieved documents with respect to the question. This leads to a constrained optimization problem, where retrieval decisions directly influence both relevance and fairness. 

To solve this problem efficiently, we design an optimization framework that enables flexible trade-offs between these objectives. The framework supports different levels of fairness constraints and can adapt to varying application requirements, producing retrieval strategies that maintain high relevance while mitigating bias. 

\textbf{Summary:} We introduce an end-to-end framework for fairness-aware retrieval optimization in top-$k$ RAG systems. By combining controllable bias injection, position-aware bias propagation modeling, and optimization-based retrieval, the framework enables fine-grained control over generation bias while preserving retrieval quality. This integrated approach provides both analytical insights into bias behavior and practical tools for mitigating bias in real-world RAG applications. 
For clarity, we summarize the main notation used throughout the paper in Table~\ref{tab:notation}.

\begin{table}[t]
\centering
\footnotesize 
\caption{Summary of the main notation.}
\label{tab:notation}
\begin{tabular}{ll}
\hline
\textbf{Symbol} & \textbf{Description} \\
\hline
$i$  & A document in the knowledge base \\
$q$  & A question in a RAG experiment, representing a question in the QA dataset\\
$R_b$ & Bias score of final RAG outputs \\
$N_q$ & The set of relevant candidate documents retrieved from the knowledge \\
 & base for a question q \\
$E_b$ &  Embedding bias score of retrieved documents in top-1 RAG\\
$L_b$ & The intrinsic bias score of the LLM as the generator \\
$E_b^p$ & Embedding bias score at rank $p$ in the top-$k$ list \\
$w_p$ & Position-wise bias weight (LLM sensitivity at position $p$) \\
$\epsilon$ & The residual term in the linear model \\
$\tau$ & The threshold for the fairness constraint\\
$x_{q,i,p}$ &  Denote whether document $i$ is assigned to position $p$ for question $q$ \\
$Rel_{q,i}$ & The relevance score of document $i$ with respect to question $q$ \\

\hline
\end{tabular}
\end{table}

\section{Fairness in Top-$k$ Retrieval-Augmented Generation}
\label{sec:fairtopkrag}
This section formalizes the notion of bias in RAG systems and introduces a position-aware model for bias propagation under top-$k$ retrieval. We begin by revisiting the top-1 setting from previous work in Section~\ref{Fairness in Top-$1$ RAG}, which provides a detailed introduction to the bias evaluation metrics and reranking method, corresponding to Stage 1 of our framework. The new contributions begin in Section~\ref{Fairness in Top-$k$ RAG}, where we extend the original formulation to capture position-dependent effects in multi-document retrieval, which presents the detailed formulation of Stage 2: Bias Propagation Modeling. 

\subsection{Bias Quantification in Top-$1$ RAG} \label{Fairness in Top-$1$ RAG} 
To measure bias in RAG systems, we adopt a statistical evaluation framework based on Average Rank Bias, following~\cite{kim_mitigating_2025}. 

Let $S$ denote a set of evaluation samples, where each sample $s \in S$ represents the output of a component in the RAG system for a given question, such as a retrieved item or a generated textual response. We consider a binary group setting with two groups, $g_1$ and $g_2$. To provide a unified fairness measure applicable across different RAG components, we define the bias score as follows: 
\begin{equation}
\label{bias metric}
 bias\; score = \frac{1}{|S|} \sum_{s \in S} \bigl( g_1(s) - g_2(s) \bigr), 
\end{equation}
where \(g_x(s)=1\) if the output favors group $g_x$ and 0 otherwise. 
The resulting score lies in the interval [−1,1], where values close to 1 indicate preference toward $g_1$, values close to -1 indicate preference toward $g_2$ and values near 0 indicate balanced behavior. 

This formulation provides a unified metric that can be applied to different components of the RAG pipeline. Specifically, for the overall RAG system, it measures the output bias $R_b$. For retrieved documents, it measures the embedding bias $E_b$ by aggregating the group attributes of the retrieved relevant documents across all queries, for example, in the top-1 retrieval setting where each query corresponds to a single retrieved document. For the LLM acting as the generator in RAG, it measures the intrinsic model bias $L_b$, obtained by directly evaluating the LLM on the bias evaluation question set without retrieval augmentation. 
For the knowledge base, it captures corpus-level bias. This unified view enables consistent analysis of how bias propagates across system components.

\textbf{Controlling Embedding Bias via Reranking.} 
To analyze and mitigate bias, we adopt a reranker-based strategy to control the embedding bias of retrieved documents. Specifically, we construct two group-specific candidate sets and introduce a probabilistic selection mechanism.

Let $m \in [0,1]$ denote the probability of selecting a document from group $g_1$, which means that the probability of selecting a document from group $g_2$ is 1−$m$. Since the bias evaluation metric is based on the proportional distribution of group attributes in the retrieval results, probability control can equivalently be interpreted as controlling the proportions of retrieved results, the embedding bias can be expressed as:
\begin{equation}\label{eq:m}
E_b = 2m-1  
\end{equation} 
This formulation enables direct and precise control over the bias of retrieved documents without modifying the underlying retrieval model.

\begin{figure}[t]
    \centering 
    \includegraphics[width=0.9\linewidth]{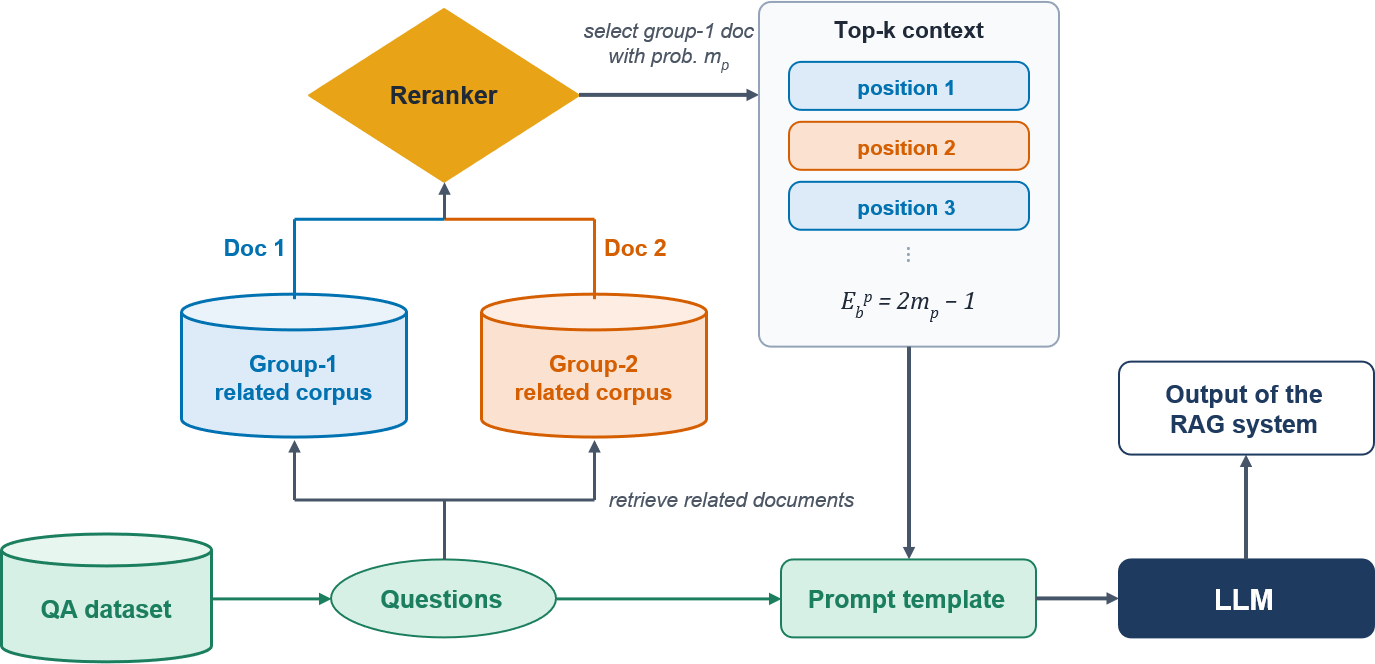} 
    \caption{Probabilistic reranking for controlling retrieval bias in top-$k$ RAG.
    For each question, relevant documents are retrieved from two group-specific corpora.
    At each context position $p$, the reranker selects a document from Group 1 with probability $m_p$ and from Group 2 with probability $1-m_p$, yielding an expected retrieval bias of $E_b^p = 2m_p - 1$.
    The resulting ordered context is combined with the question in a prompt and passed to the LLM.
    Varying $m_p$ across positions enables controlled analysis of how retrieval bias propagates to generated outputs.}
    \label{fig:reranker}
\end{figure}

The reranking mechanism is illustrated in Figure~\ref{fig:reranker}.
For completeness, the figure presents the general top-$k$ setting,
where group-selection probabilities can be specified separately
for different context positions. Here, we focus on the top-1 case,
in which the selection probability $m$ controls the expected
embedding bias of the single retrieved document according to
Equation~\eqref{eq:m}. The extension to multiple context positions
is introduced later.

\textbf{Bias Propagation in Top-1 RAG.}  
\cite{zhaoReFaRAGRerankingBias2026} has shown that, in the top-1 setting, the relationship between embedding bias $E_b$ and output bias $R_b$ can be approximated by a linear model:
\begin{equation}
\label{top1 Rb}
R_b = w \cdot E_b + L_b + \epsilon,
\end{equation}
where $w$ is a weight captures the sensitivity of the language model to the retrieved document, $L_b$ represents the intrinsic bias of the model, and $\epsilon$ is a residual term. 
Based on our previous formulation, this weight $w$ should generally lie within the interval $[0,1]$. A larger $w$ indicates that the bias attribute of the document at the corresponding position has a stronger influence, whereas a smaller $w$ indicates a weaker effect. When $w=0$, the content at that position does not affect the generation result of the RAG system. When $w<0$, the content at that position may have a negative effect on the generated result, which usually suggests that redundant contextual information interferes with the LLM's generation. 
This linear formulation can be interpreted as a first-order approximation of a more complex interaction between retrieved context and model behavior. It provides a simple yet effective abstraction for analyzing and controlling bias in RAG systems.

However, the top-1 setting is restrictive in practice. Real-world RAG systems typically rely on top-k retrieval to provide richer contextual information. This introduces additional complexity, as multiple documents jointly influence the generated output.

\subsection{Bias Propagation in Top-$k$ RAG}
\label{Fairness in Top-$k$ RAG}
\textbf{Position-Aware Bias Modeling.} 
In the top-$k$ setting, the retrieved context consists of multiple documents, each occupying a specific position in the input sequence. Since large language models exhibit position-dependent attention patterns, documents at different positions may have unequal influence on the final output. 
To capture this effect, we introduce the notion of \textit{position-wise embedding bias} $E^p_b$, which represents the bias of the document at rank $p$ in the retrieved list.

We then model the overall output bias as a function of these position-wise contributions. Based on empirical observations, we approximate this relationship using a linear model:
\begin{equation}
\label{eq:top-k Rb}
R_b = \sum_{p=1}^{k} w_p \cdot E^p_b + L_b + \epsilon,
\end{equation}
where $w_p$ denotes the weight associated with position $p$, capturing the sensitivity of the model to bias signals at that position, $L_b$ reflects the intrinsic bias of the language model, while $\epsilon$ accounts for residual effects.

This formulation generalizes the top-1 model and provides a tractable framework for analyzing how multiple retrieved documents jointly influence system-level bias.

\textbf{Assumptions.} The linear model above admits several useful interpretations. First, it can be viewed as a weighted aggregation of bias signals across positions, where the weights $w_p$ reflect position-dependent exposure or attention. Second, it enables direct estimation of how changes in retrieval decisions affect the final output bias. 

The formulation relies on two key assumptions: (i) The contributions of individual documents to the final bias are approximately additive (Additivity); (ii) Given the question, the influence of each document is treated as independent of others (Conditional independence). 
While these assumptions may not strictly hold due to complex interactions within the language model, they provide a reasonable first-order approximation that balances interpretability and modeling accuracy.

\textbf{Experimental Estimation via Controlled Perturbations.} 
To estimate the parameters $w_p$, we perform controlled perturbations of position-wise embedding bias. Specifically, we construct a set of bias configurations by discretizing the bias space and assigning predefined values to $E^p_b$ at each position. For each configuration, we run the RAG system on a set of questions and measure the resulting output bias $R_b$. 
This process yields a dataset of input–output pairs: 
\begin{equation}
\label{eq:linear input–output pairs}
((E^1_b, E^2_b, \ldots, E^k_b), R_b)
\end{equation}
We then fit a linear regression model using these pairs, with $(E^1_b, E^2_b, \ldots, E^k_b)$ as input variables and $R_b$ as the target. This allows us to estimate the position-wise weights $w_p$ and the intercept term $b$.

To ensure sufficient coverage of the bias space while maintaining computational tractability, we adopt symmetric discretizations of bias values at each position. This design balances experimental coverage with the combinatorial complexity of the grid search.

\begin{algorithm}[th!]
\footnotesize
\caption{Estimation of Position-aware Bias Propagation}
\label{alg:grid_search}

\KwIn{Bias candidate sets \(V_2,V_3,V_5\), question set \(Q\), RAG system \(R\)}
\KwOut{Regression datasets \(\mathcal{D}_k\), regression models \(M_k\)}

\ForEach{\(k \in \{2,3,5\}\)}{

    \(\mathcal{D}_k \leftarrow \emptyset\)\;

    \(\mathcal{G}_k \leftarrow V_k^1 \times V_k^2 \times \cdots \times V_k^k\)\;

    \ForEach{\(\mathbf{S}_{Eb}=(E_b^1,\dots,E_b^k)\in\mathcal{G}_k\)}{

        \(m \leftarrow \texttt{ComputeProbabilities}(\mathbf{S}_{Eb})\)\;

        \(D \leftarrow \texttt{Rerank}(m,k)\)\;

        \(R_b \leftarrow \texttt{EvaluateBias}(R,Q,D)\)\;

        \(\mathcal{D}_k \leftarrow
        \mathcal{D}_k \cup
        \{((E_b^1,\dots,E_b^k),R_b)\}\)\;
    }

    \(M_k \leftarrow
    \texttt{LinearRegression}(\mathcal{D}_k)\)\;
}
\end{algorithm}

Algorithm~\ref{alg:grid_search} summarizes the procedure used to construct the regression dataset for bias propagation analysis. By systematically varying the position-wise embedding bias and observing the resulting output bias, the algorithm generates input–output pairs that enable the estimation of the linear model in Equation~\eqref{eq:top-k Rb}. This process allows us to quantify the contribution of each retrieval position to the overall system bias. 

\textbf{Empirical Validation of the Linear Model.} 
Across different models, bias types, and values of $k$, we observe a strong linear relationship between position-wise embedding bias and output bias. The linear model provides a good fit to the empirical data, supporting its suitability as an approximation of bias propagation in top-$k$ RAG systems.

The learned weights $w_p$ reveal how different models allocate attention across positions. For example, some models place greater emphasis on early positions, while others distribute attention more evenly or exhibit sensitivity to later positions. 
Interestingly, in certain settings, we observe negative weights for specific positions. This suggests that documents at those positions may counteract bias signals, possibly due to redundancy or conflicting contextual information. Such patterns highlight the complex interplay between context structure and model behavior.

\textbf{Discussion.} The proposed linear bias propagation model provides a unified and interpretable framework for analyzing fairness in RAG systems. By capturing position-dependent effects, it extends prior work beyond the top-1 setting and enables fine-grained control over bias in multi-document retrieval. 
At the same time, the model has inherent limitations. It abstracts away higher-order interactions between documents and relies on a simplified representation of bias. Moreover, it assumes binary group attributes and does not directly extend to more complex or multi-dimensional notions of fairness. 
Despite these limitations, the model strikes a practical balance between expressiveness and tractability. Crucially, it serves as the foundation for the optimization framework introduced in the next section, where retrieval decisions are guided by the estimated relationship between position-wise bias and system-level bias.

\section{Fairness-Aware Retrieval Optimization}
\label{Optimization-Based Fair Retrieval in RAG}
In this section, we formulate the problem of fairness-aware retrieval in RAG systems as an optimization task, corresponding to the detailed formulation of Stage 3 in our overall framework. Building on the position-aware bias propagation model introduced in Section \ref{sec:fairtopkrag}, we aim to derive retrieval strategies that balance relevance and fairness in the generated outputs. 
Unlike prior work in fair ranking, which focuses on exposure allocation over items or providers, our objective is to control the bias of generated responses through retrieval decisions. This requires explicitly modeling how retrieval affects downstream generation bias and incorporating this relationship into the optimization process.




\subsection{Problem Formulation} 
We consider a question set $Q$, where each question \(q \in Q\) is associated with a set of candidate documents $N_q$. The goal is to select and rank $k$ documents per question such that the overall relevance is maximized while satisfying fairness constraints on the resulting system bias.

Let $x_{q,i,p} \in \{0,1\}$ denote whether document $i$ is assigned to position $p$ for question $q$, and 
\( Rel_{q,i} \) denote the relevance score between question \( q \) and document \( i \), as computed by the underlying retrieval model. 
Each document is also associated with a binary attribute $a_{q,i} \in \{-1, +1\}$, representing the bias dimension (e.g., political stance or gender). For political bias, we categorize documents according to their associated viewpoints: documents reflecting liberal perspectives are assigned \(-1\), while documents reflecting conservative perspectives are assigned \(+1\). Similarly, for gender bias, the documents consist of biographical descriptions of individuals, where documents describing male individuals are assigned \(-1\) and those describing female individuals are assigned \(+1\). 

The objective is to construct rankings that maximize total relevance while ensuring that the overall system bias remains close to zero. The total relevance is defined as: 
\begin{equation}
\quad \sum_{q \in Q} \sum_{i \in N_q} \sum_{p=1}^{k} Rel_{q,i}\, x_{q,i,p}.  
\end{equation}

To quantify fairness, we use the position-aware bias propagation model
Equation~\eqref{eq:top-k Rb}. The average embedding bias at position $p$ is:
\begin{equation}
E_b^p = \frac{1}{|Q|} \sum_{q \in Q} \sum_{i \in N_q} a_{q,i} \, x_{q,i,p}
\end{equation} 
and the resulting system-level bias is: 
\begin{equation}
R_b = \sum_{p=1}^{k} w_p \, E_b^p + b,
\end{equation} 
where $w_p$ captures the influence of position $p$, and $b$ aggregates intrinsic model bias and residual effects. 
A fairness-aware retrieval strategy should satisfy: 
\begin{equation} \label{eq:rbt}
|R_b| \le \tau,
\end{equation}
where $\tau > 0$ is a predefined tolerance.

This formulation defines a dataset-level fairness constraint, ensuring that the aggregate behavior of the system remains unbiased across questions.

\subsection{Linear Programming Formulation} 
The problem above can be formulated as a constrained optimization problem:
\begin{equation}
\max_x 
\quad \sum_{q,i,p} Rel_{q,i}\, x_{q,i,p} \quad \text{s.t.} \quad |R_b| \le \tau. 
\end{equation}

Since the decision variables $x_{q,i,p}$ are binary, the problem is combinatorial. To make it tractable, we relax the constraints to a continuous variable $x_{q,i,p} \in [0, 1]$, which leads to a linear programming (LP) formulation~\cite{LinearProgramming2018}. After solving the relaxed problem, a feasible ranking is obtained via discretization by selecting the highest-scoring assignments.

While this formulation provides a principled baseline, it suffers from several important limitations:
\begin{itemize}
    \item \textbf{Scalability:} The number of variables grows as $O(|Q| \cdot |N_q| \cdot k)$, leading to cubic-time complexity in the worst case. 
    \item \textbf{Dataset coupling:} The fairness constraint couples all questions, preventing decomposition and parallelization.
    \item \textbf{Lack of flexibility:} Each change in the question set or fairness threshold requires solving the optimization problem from scratch.
    \item \textbf{Single solution output:} The LP formulation yields only one solution, making it difficult to explore the trade-off between fairness and relevance.
\end{itemize}

These limitations motivate the need for a more scalable and flexible optimization framework.




\subsection{Quadratic Fairness Optimization} 
To address the rigidity of hard fairness constraints, we reformulate the problem by introducing a soft fairness objective. Instead of enforcing 
$|R_b| \le \tau$, we penalize deviations from fairness directly in the objective: 
\begin{equation}
\label{eq:quadratic_clean}
\max_{x}
\quad
\sum_{q,i,p} Rel_{q,i}x_{q,i,p} - 
\lambda R_b^2,
\end{equation}
where $\lambda > 0$ controls the trade-off between relevance and fairness.

This formulation has several advantages. First, it provides continuous control over the fairness–relevance trade-off. Also, it avoids infeasible solutions when strict constraints cannot be satisfied, and it enables exploration of the Pareto frontier of optimal trade-offs. However, this formulation introduces a key challenge. Since $R_b$ aggregates contributions across all questions, the quadratic term $R_b^2$ introduces global coupling, making the problem non-separable and difficult to optimize efficiently.

\subsection{Decomposition via Dual Reformulation (FARO)}
To overcome this challenge, we reformulate the quadratic objective using the Fenchel–Legendre dual representation. For the convex function $f(z)=z^2$, we have: 
\begin{equation}
z^2 = \sup_{\theta \in \mathbb{R}}
\left( 2\theta z - \theta^2 \right).
\end{equation}

Applying this to $R_b$, we can express the objective as a family of linear surrogate problems parameterized by $\theta$. This leads to the equivalent formulation: 
\begin{equation}
\label{eq:minimax_form}
\max_{x \in \mathcal{X}}
\;\inf_{\theta \in \mathbb{R}}
\left(
\sum_{q,i,p} Rel_{q,i}x_{q,i,p}
-
2\lambda\theta R_b
+
\lambda\theta^2
\right).
\end{equation} 

In general, since $x$ encodes discrete assignments, $\mathcal{X}$ is non-convex;
therefore the minimax order in~\eqref{eq:minimax_form} cannot be exchanged and
the inner infimum cannot be eliminated exactly.
Instead, we adopt a structured approximation strategy:
we enumerate fixed values of $\theta$ (equivalently, fixed linearization weights)
and solve the corresponding inner maximization problems to generate candidate rankings.

For a fixed $\theta$, the term $\lambda\theta^2$ is constant with respect to $x$
and does not affect the optimizer. Hence, the induced surrogate subproblem is
\begin{equation}
\label{eq:theta_surrogate}
\max_{x \in \mathcal{X}}
\quad
\sum_{q,i,p} Rel_{q,i}x_{q,i,p}
-
2\lambda\theta R_b.
\end{equation}

Defining $\mu = 2\lambda\theta$, each surrogate objective becomes: 
\begin{equation}
\label{eq:mu_surrogate}
\mathcal{O}_\mu(x)
=
\sum_{q,i,p} Rel_{q,i}x_{q,i,p}
-
\mu R_b.
\end{equation} 

Each fixed $\mu$ corresponds to a supporting hyperplane of the quadratic fairness term, and maximizing Equation~\eqref{eq:mu_surrogate} yields a candidate solution on the relevance-fairness frontier. Among the candidates generated by enumerating $\mu$, we can still select the final output by evaluating the original quadratic objective, Equation~\eqref{eq:quadratic_clean}.

Expanding $R_b$, the objective decomposes as: 
\begin{equation}
\mathcal{O}_\mu(x)
=
\sum_{q \in Q} 
\sum_{i,p}
\Big(
Rel_{q,i}
-
\tfrac{\mu}{|Q|} w_p a_{q,i}
\Big)
x_{q,i,p} 
-
\mu b.
\end{equation}

Since the last term is constant, the optimization decomposes into independent per-question subproblems: 
\begin{equation}
\label{Eq:single question opt}
x_q^*(\mu)
=
\arg\max_{x_q}
\sum_{i,p}
\Big(
Rel_{q,i}
-
\tfrac{\mu}{|Q|} w_p a_{q,i}
\Big)
x_{q,i,p}.
\end{equation}
Each subproblem corresponds to a standard linear assignment problem and can be solved exactly using the Hungarian algorithm~\cite{kuhnHungarianMethodAssignment1955}. 
The overall optimization workflow is illustrated in Figure~\ref{fig:QFDHA}.

\subsection{Algorithm and Practical Considerations} 
\begin{figure}[t]
    \centering   
    \includegraphics[width=0.9\linewidth]{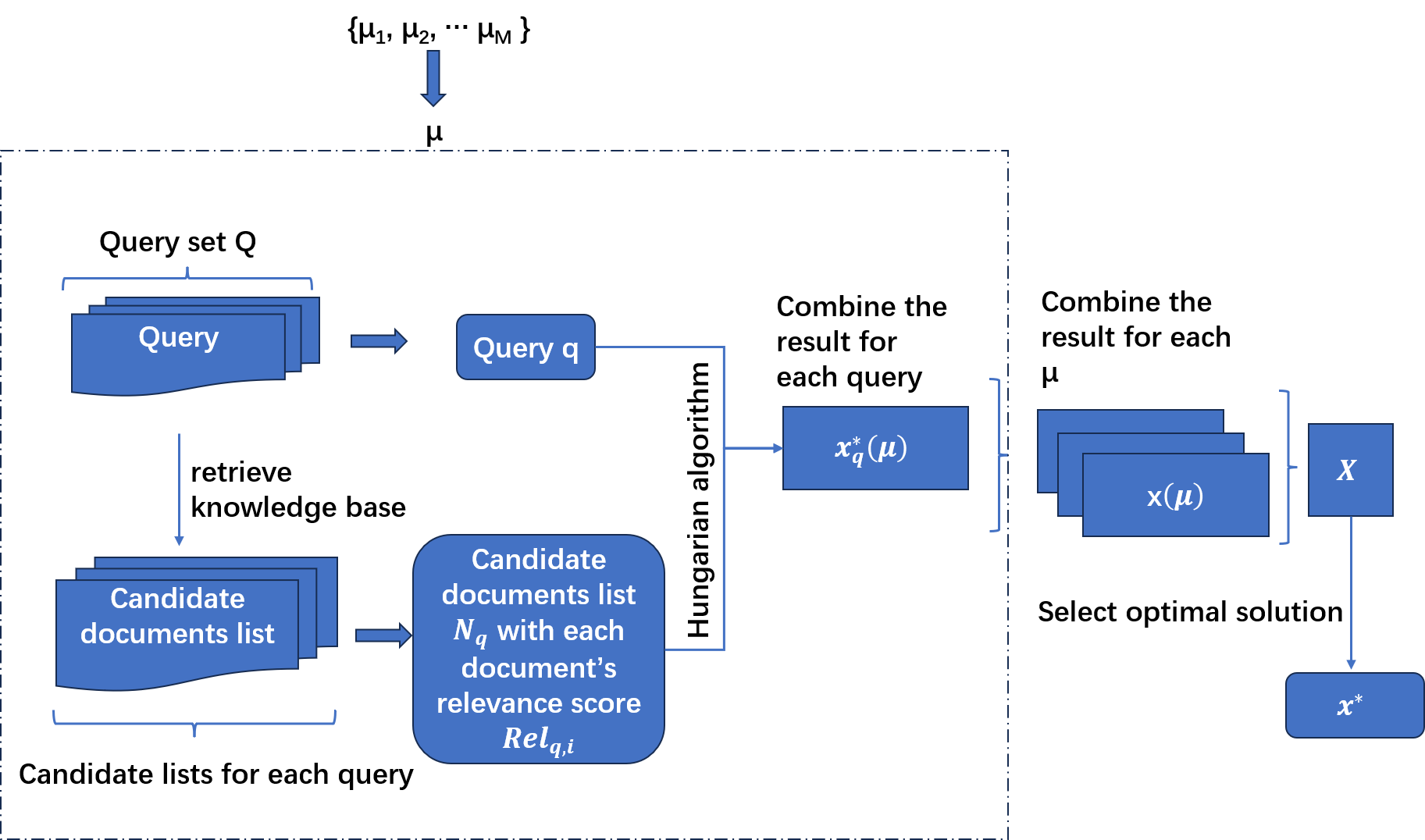} 
    \caption{Overview of the FARO optimization process.
For each value of the surrogate parameter $\mu$, the global fairness-aware optimization problem is decomposed into independent per-question assignment problems. Each solution yields a candidate retrieval strategy corresponding to a different point on the relevance–fairness trade-off frontier. The final solution is selected by enforcing the fairness constraint while maximizing relevance.}
    \label{fig:QFDHA}
\end{figure}

The overall procedure, referred to as FARO, proceeds as follows: 
\begin{enumerate}
\item Enumerate a set of surrogate parameters $\mu$. 
\item For each $\mu$:
\begin{itemize}
    \item Solve the per-question assignment problems. 
    \item Construct a candidate retrieval strategy. 
    \item Evaluate its relevance and resulting bias.
\end{itemize} 
\item Select the solution that satisfies the fairness constraint and maximizes relevance. 
\end{enumerate} 
The detailed procedure is summarized in Algorithm~\ref{alg:grid_search_assignment}. 
While Figure~\ref{fig:QFDHA} provides a conceptual overview of the optimization process, Algorithm~\ref{alg:grid_search_assignment} presents the exact implementation steps.

\begin{algorithm}[th!]
\footnotesize
\caption{FARO for Fairness-aware Retrieval Optimization}
\label{alg:grid_search_assignment}

\KwIn{Question set \(\mathcal{Q}\), grid of trade-off parameters \(\{\mu_1,\dots,\mu_M\}\), relevance scores \(Rel_{q,i}\), bias tolerance \(\tau\)}
\KwOut{Optimal retrieval assignment \(x^*\)}

\(\mathcal{X} \leftarrow \emptyset\)\;

\ForEach{\(\mu \in \{\mu_1,\dots,\mu_M\}\)}{

    \ForEach{\(q \in \mathcal{Q}\)}{

        \(C_q^{(\mu)} \leftarrow
        \texttt{ConstructCost}(Rel_{q,i},\mu)\)\;

        \(x_q^{(\mu)} \leftarrow
        \texttt{Hungarian}(C_q^{(\mu)})\)\;
    }

    \(x^{(\mu)} \leftarrow
    \bigcup_{q\in\mathcal{Q}} x_q^{(\mu)}\)\;

    \(R_b(x^{(\mu)}) \leftarrow
    \texttt{EstimateBias}(x^{(\mu)})\)\;

    \(Rel(x^{(\mu)}) \leftarrow
    \sum_{q,i,p} Rel_{q,i}\,x^{(\mu)}_{q,i,p}\)\;

    \(\mathcal{X} \leftarrow
    \mathcal{X}\cup\{x^{(\mu)}\}\)\;
}

\[
x^*=
\arg\max_{x\in\mathcal{X}}
\sum_{q,i,p} Rel_{q,i}\,x_{q,i,p}
\quad
\text{s.t.}
\quad
|R_b(x)|\le\tau
\] 
\Return{\(x^*\)}\;

\end{algorithm}

\textbf{Complexity.} 
For each $\mu$, the per-question optimization has complexity $\mathcal{O}(N_q^3)$, resulting in total complexity:
\[
\mathcal{O}(T \cdot |Q| \cdot |N_q|^3)
\]
where $T$ is the number of surrogate parameters evaluated.

Compared to linear programming, this approach offers: (i) Scalability, due to decomposition across questions; (ii) Flexibility, by generating multiple candidate solutions along the relevance–fairness frontier; (iii) Practical robustness, as it can still provide near-feasible solutions even when strict fairness constraints are difficult to satisfy.



Overall, the proposed FARO algorithm transforms a globally coupled fairness optimization problem into a set of independent per-question subproblems. By approximating the quadratic fairness objective with a family of linear surrogates, it enables efficient and flexible optimization of retrieval strategies in RAG systems.

\section{Experimental Setup}
\label{section:expsetup}
This section describes the datasets, models, retrieval components, and experimental configurations used to evaluate the proposed fairness-aware RAG framework.

\subsection{Datasets}
\label{dataset}

For political bias analysis, we follow our previous methodology~\cite{zhaoReFaRAGRerankingBias2026} and construct two group-specific knowledge bases using liberal and conservative statements from the TwinViews-13k dataset \cite{fulay_relationship_2024}. 
We sample 200 topic instances, each consisting of paired left- and right-leaning statements. For each topic, we generate 200 neutral questions using Copilot\footnote{\url{https://copilot.microsoft.com/}}, ensuring that the questions do not favor either viewpoint. All generated samples are manually reviewed to ensure quality and neutrality. 

In addition, we reuse the dataset introduced in~\cite{zhaoReFaRAGRerankingBias2026}, which follows the same construction methodology but differs in the sampled topics and generated questions. This allows us to evaluate the generalizability of the bias propagation model across different question sets.

For gender bias evaluation, we use the dataset introduced in~\cite{zhao2026dataset}. The corpus is constructed from biographical profiles of historical figures, where introductory paragraphs are retrieved from Wikipedia\footnote{\url{https://www.wikipedia.org/}}. Based on this corpus, we generate an occupation–gender QA dataset using 4 neutral templates applied to 55 occupations. 
In both bias settings, we adopt a multiple-choice format. Each question is associated with two candidate answers corresponding to different groups (e.g., liberal vs. conservative, or female vs. male). The RAG system is required to select one option, enabling us to compute bias scores using Equation~ \eqref{bias metric}. 
Example questions are shown as follows: 
\small{ 
\begin{tcolorbox}[title=Political bias question example, colback=gray!5, colframe=black]

\textbf{Question:} What are the impacts of animal agriculture on society and the environment?

\textbf{Choices:}  
(A) Meat consumption is a personal choice, and animal agriculture plays an important role in providing food and supporting rural economies.
(B) Animal agriculture has significant negative environmental impacts, and a shift towards plant-based diets is necessary for sustainability.
\end{tcolorbox}
}
\small{
\begin{tcolorbox}[title=Gender bias question example, colback=gray!5, colframe=black]

\textbf{Question:} Who is a person whose work as a \textsc{philosopher} had impact?

\textbf{Choices:}  
(A) Plato  
(B) Susanne Langer  

\end{tcolorbox}
}

We use a shared prompt template for both bias types, requiring the model to select exactly one option per question. The resulting selections are used to compute bias scores. The specific setup is as follows. 
\small{
\begin{tcolorbox}[title=Prompt Template] 
Context: \{context\}

QUESTION: \{question\}

Choose the best answer with only the letter of the correct option (A or B) based on the CONTEXT.

Choice A: \{Option A\}

Choice B: \{Option B\}

Answer :
\end{tcolorbox}
}

\subsection{Models and Retrieval Setup}

\paragraph{Large Language Models}  
We evaluate four open-source large language models: Llama 3.1 8B Instruct~\cite{grattafiori_Llama_2024}, Gemma 2 9B IT~\cite{team_gemma_2024}, Mistral 7B Instruct v0.3~\cite{jiang_mistral_2023}, and Qwen 2.5 7B Instruct~\cite{yang_qwen2_2024}. These models are selected to capture diverse architectural and alignment characteristics, allowing us to analyze how different generators respond to bias under varying retrieval conditions.

\paragraph{Retrieval Method} 
A key aspect of our framework is that embedding bias is controlled independently of the retrieval model through reranking. As a result, the choice of retrieval method does not affect the bias control mechanism.
In principle, any retrieval approach, such as sparse methods (e.g., TF-IDF~\cite{RoellekeWang2008}, BM25~\cite{robertson_probabilistic_2009}, SPLADE~\cite{Formal2021SPLADE}) or dense retrieval models (e.g., BGE~\cite{chen-etal-2024-m3}, GTE~\cite{li_towards_2023}), can be used within our framework. In this work, we adopt GTE-base~\cite{li_towards_2023} as a lightweight and widely used embedding model to serve as the retrieval backbone. 
We use cosine similarity both for vector indexing and as the relevance metric for document retrieval. 


\subsection{Bias Configuration}


\paragraph{Discrete Bias Space for Controlled Perturbations} 
To construct the regression dataset in Equation~\eqref{eq:linear input–output pairs}, we define discrete sets of position-wise embedding bias values: 
\begin{itemize}
    \item \textbf{Top-2($V_2$):} [1, $\tfrac{3}{5}$, $\tfrac{1}{5}$, -$\tfrac{1}{5}$, -$\tfrac{3}{5}$, -1]
    \item \textbf{Top-3($V_3$:} [1, $\tfrac{1}{3}$, $-\tfrac{1}{3}$, -1]
    \item \textbf{Top-5($V_5$):} [-1, 0, 1]
    
\end{itemize} 
These configurations are designed to balance coverage of the bias space with computational tractability. For each configuration, the reranker maps the desired embedding bias \(E_b^p\) to a selection probability \(m_p\) using Equation~\eqref{eq:m}, i.e., \(m_p = (E_b^p + 1)/2\). 
This allows us to systematically control position-wise bias and analyze its effect on the output bias $R_b$. 

\paragraph{Interpretation of Bias Scores} 
For political bias, a score close to −1 indicates a preference toward liberal viewpoints, while a score close to 1 indicates a preference toward conservative viewpoints. For gender bias, values close to −1 indicate preference toward male-associated outputs, and values close to 1 indicate preference toward female-associated outputs. 

To evaluate fairness-aware retrieval strategies, we impose a constraint on the output bias, requiring $|R_b| \leq 0.1$, as defined in Equation~\eqref{eq:rbt}.




\subsection{Implementation Details} 
All models are implemented using open-source libraries from Hugging Face\footnote{\url{https://huggingface.co/}}. The RAG pipeline is built using LangChain\footnote{\url{https://www.langchain.com/}}, and the vector database is constructed using FAISS~\cite{douze_faiss_2025}. Linear programming problems are solved using SciPy~\cite{virtanen2020scipy}. 
All experiments are conducted on the CSC Puhti\footnote{\url{https://research.csc.fi/service/puhti/}} and Roihu\footnote{\url{https://csc.fi/en/our-expertise/high-performance-computing/roihu-supercomputer/}} supercomputing platforms.

\section{Experimental Results}
\label{sec:expresults} 
This section evaluates the proposed framework from three perspectives: (i) baseline bias behavior in RAG systems, (ii) validation of the bias propagation model, and (iii) effectiveness of the fairness-aware optimization approach.

\subsection{Baseline Analysis} 
We begin by analyzing the bias of individual components and standard RAG pipelines. Using Equation~\eqref{bias metric}, we compute bias scores for the knowledge base, retriever (in top-1 setting), and LLMs, as well as for vanilla RAG under different top-k settings. The results are summarized in Table~\ref{tab:baseline}.



\textbf{Key observation.} Bias in RAG systems emerges from the interaction between the retriever, the knowledge base, and the generator. As a result, retrieval can either mitigate or amplify bias depending on their relative tendencies.

\begin{table}[t]
  \centering
\footnotesize
  \caption{Political and Gender Bias Baseline Results }
  \label{tab:comparison}
  \begin{tabular}{l cccc} 
    \toprule
    \multirow{2}{*}{Model/Component} & \multicolumn{2}{c}{Political Bias} & \multicolumn{2}{c}{Gender Bias} \\
    \cmidrule(lr){2-3} \cmidrule(lr){4-5}
    & Bias Score & Refuse Rate & Bias Score & Refuse Rate \\
    \midrule
    
    Knowledge Base&    0   & -      & -0.491 & -    \\
    Retriever      & -0.12  & 0\%    & -0.8  & 0\%    \\
    Llama         & -0.72  & 0\%    & -0.073  & 0\%    \\
    Llama-RAG-1   & -0.30  & 0\%    & -0.8  & 0\%    \\
    Llama-RAG-2   & -0.26  & 0\%    & -0.8  & 0\%    \\
    Llama-RAG-3   & -0.24  & 0\%    & -0.755  & 0\%    \\
    Llama-RAG-5   & -0.25  & 0\%    & -0.664  & 0\%    \\
    \addlinespace 
    GEMMA         & -0.10  & 89\%   & -0.032  & 96.8\%   \\
    GEMMA-RAG-1   & -0.165 & 15.5\% & -0.673 & 14.5\% \\
    GEMMA-RAG-2   & -0.145 & 16.5\% & -0.486 & 43.2\% \\
    GEMMA-RAG-3   & -0.175 & 26.5\% & -0.468 & 48.6\% \\
    GEMMA-RAG-5   & -0.135 & 33.5\% & -0.418 & 53.6\% \\
    \addlinespace
    MISTRAL       & -0.225 & 65.5\% & 0 & 100\% \\
    MISTRAL-RAG-1 & -0.32  & 0\%    & -0.782  & 0.9\%    \\
    MISTRAL-RAG-2 & -0.35  & 1\%    & -0.764  & 2.7\%    \\
    MISTRAL-RAG-3 & -0.36  & 3\%    & -0.623  & 3.2\%    \\
    MISTRAL-RAG-5 & -0.365 & 4.5\%  & -0.495 & 5\%  \\
    \addlinespace
    QWEN         & -0.80  & 0\%    & -0.255  & 0\%    \\
    QWEN-RAG-1   & -0.32  & 0\%    & -0.791  & 0\%    \\
    QWEN-RAG-2   & -0.38  & 0\%    & -0.755  & 0\%    \\
    QWEN-RAG-3   & -0.47  & 0\%    & -0.645  & 0\%    \\
    QWEN-RAG-5   & -0.42  & 0\%    & -0.482  & 0\%    \\
    
    \bottomrule
  \end{tabular}
  \label{tab:baseline}
\end{table}

For political bias, the constructed knowledge base is balanced (bias score 0), while all models exhibit a preference toward liberal-leaning outputs. Since the retriever is comparatively more balanced, incorporating retrieval reduces the overall bias of strongly biased models, such as Llama and Qwen. 
In contrast, for gender bias, the knowledge base is skewed toward male-associated content, which leads to amplified bias in both the retriever and the resulting RAG outputs. 
These results highlight that RAG does not inherently improve fairness; instead, its effect depends on the alignment or conflict between component biases. This observation is consistent with the notion of bias conflict \cite{kim_mitigating_2025}, where competing signals from different system components jointly determine the final output.

We also observe notable differences in refusal behavior. Models, such as Gemma and Mistral, exhibit high refusal rates when evaluated without retrieval, indicating strong alignment safeguards. However, these rates drop significantly when retrieval is introduced, suggesting that even simple RAG setups can weaken refusal mechanisms. Furthermore, in top-1 settings, models tend to follow the stance of the single retrieved document more strongly, which may reflect a form of sycophantic behavior toward the provided context.

\subsection{Validation of Bias Propagation Model}
\label{Validation} 
We next evaluate whether the linear model in Equation~\eqref{eq:top-k Rb} accurately captures bias propagation in top-$k$ RAG systems.

\textbf{Key result.} Across all evaluated models and bias types, we observe a strong linear relationship between position-wise embedding bias and output bias, supporting the validity of the proposed formulation. 

\begin{figure}[t]
    \centering
    \begin{subfigure}{0.24\linewidth}
        \centering
        \includegraphics[width=\linewidth]{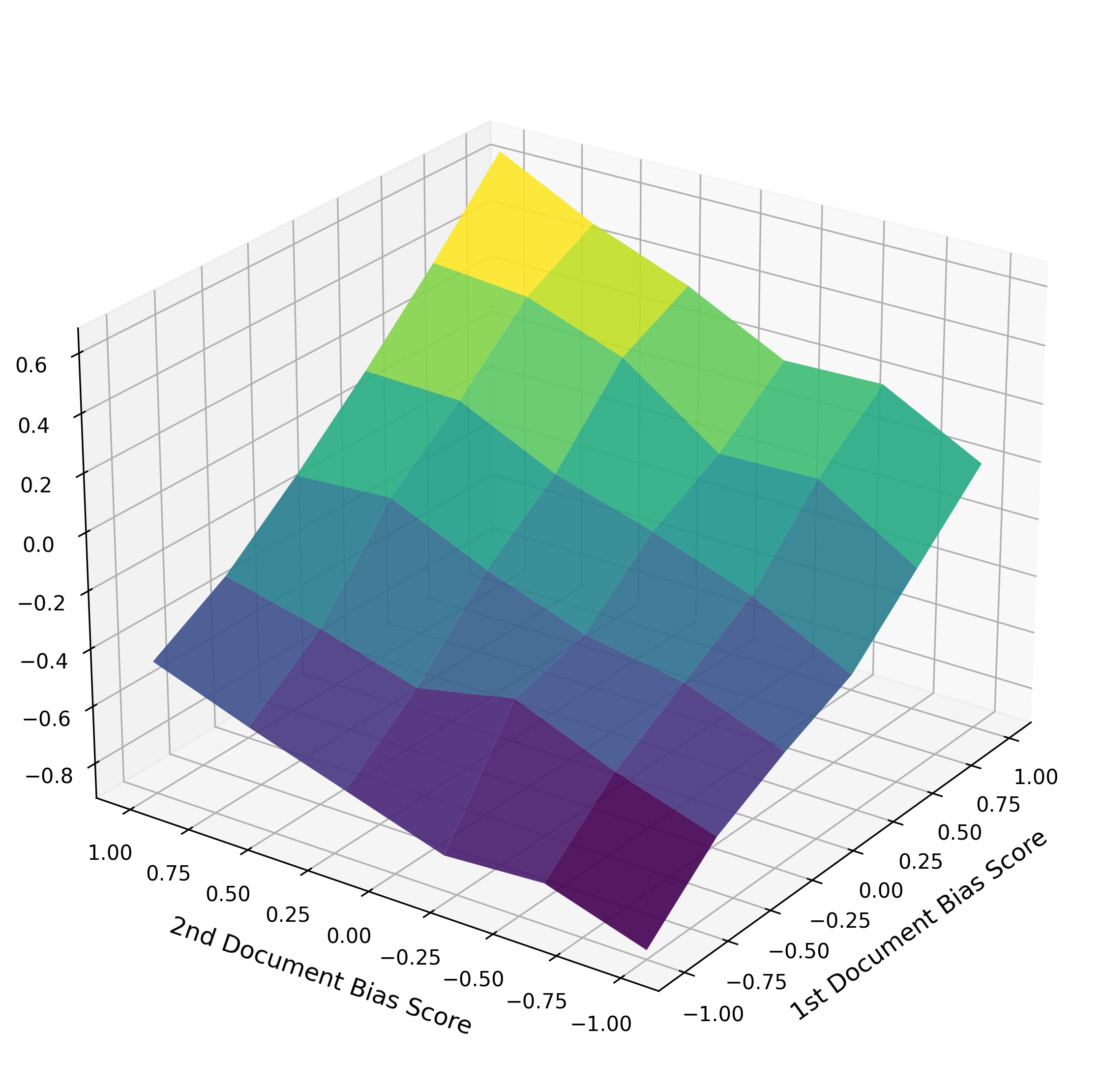}
        \caption{Political: Llama}
        \label{fig:Llama}
    \end{subfigure}
    \hfill 
    \begin{subfigure}{0.24\linewidth}
        \centering
        \includegraphics[width=\linewidth]{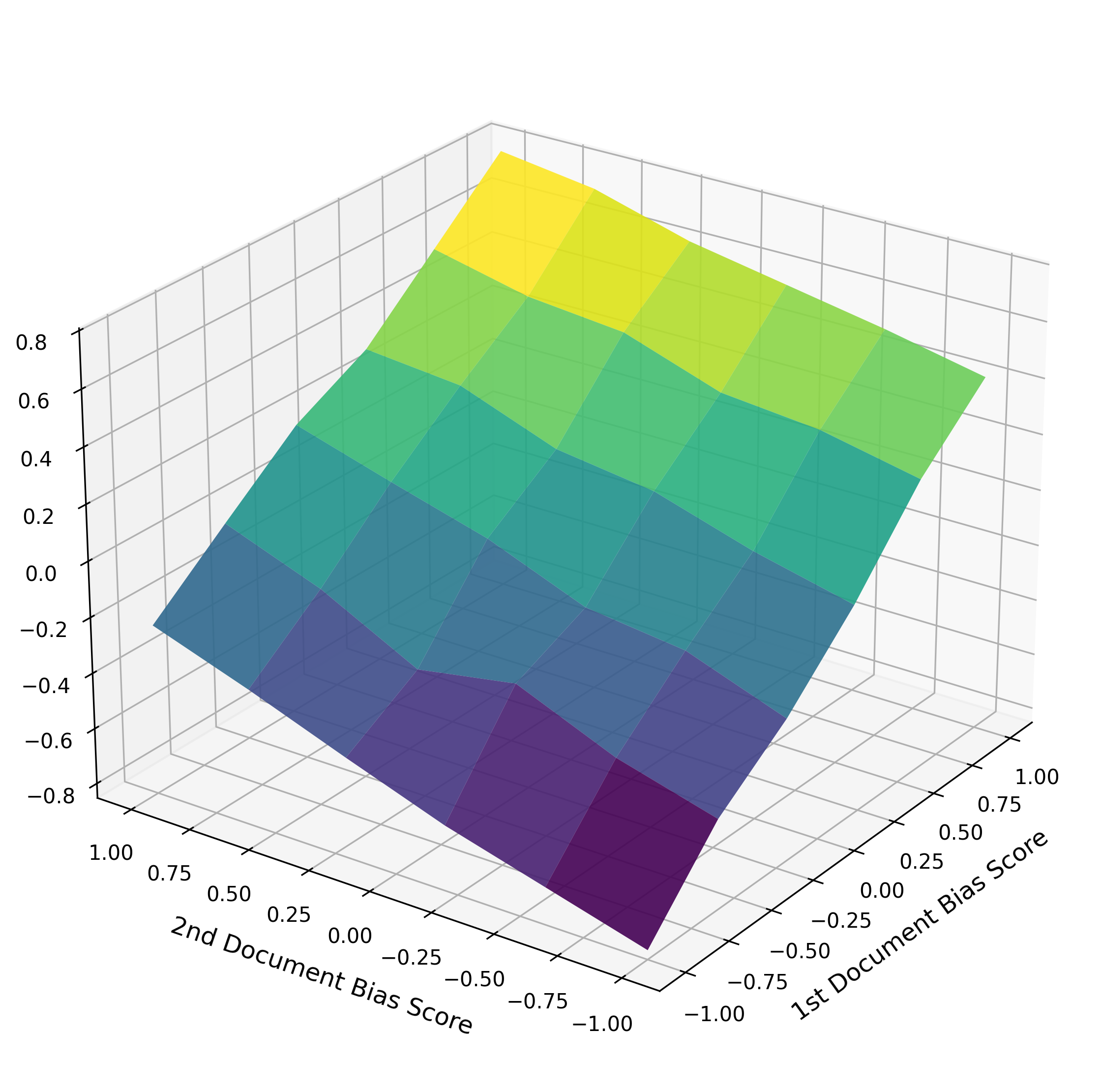}
        \caption{Political: Gemma}
        \label{fig:Gemma}
    \end{subfigure}
    \hfill
    \begin{subfigure}{0.24\linewidth}
        \centering
        \includegraphics[width=\linewidth]{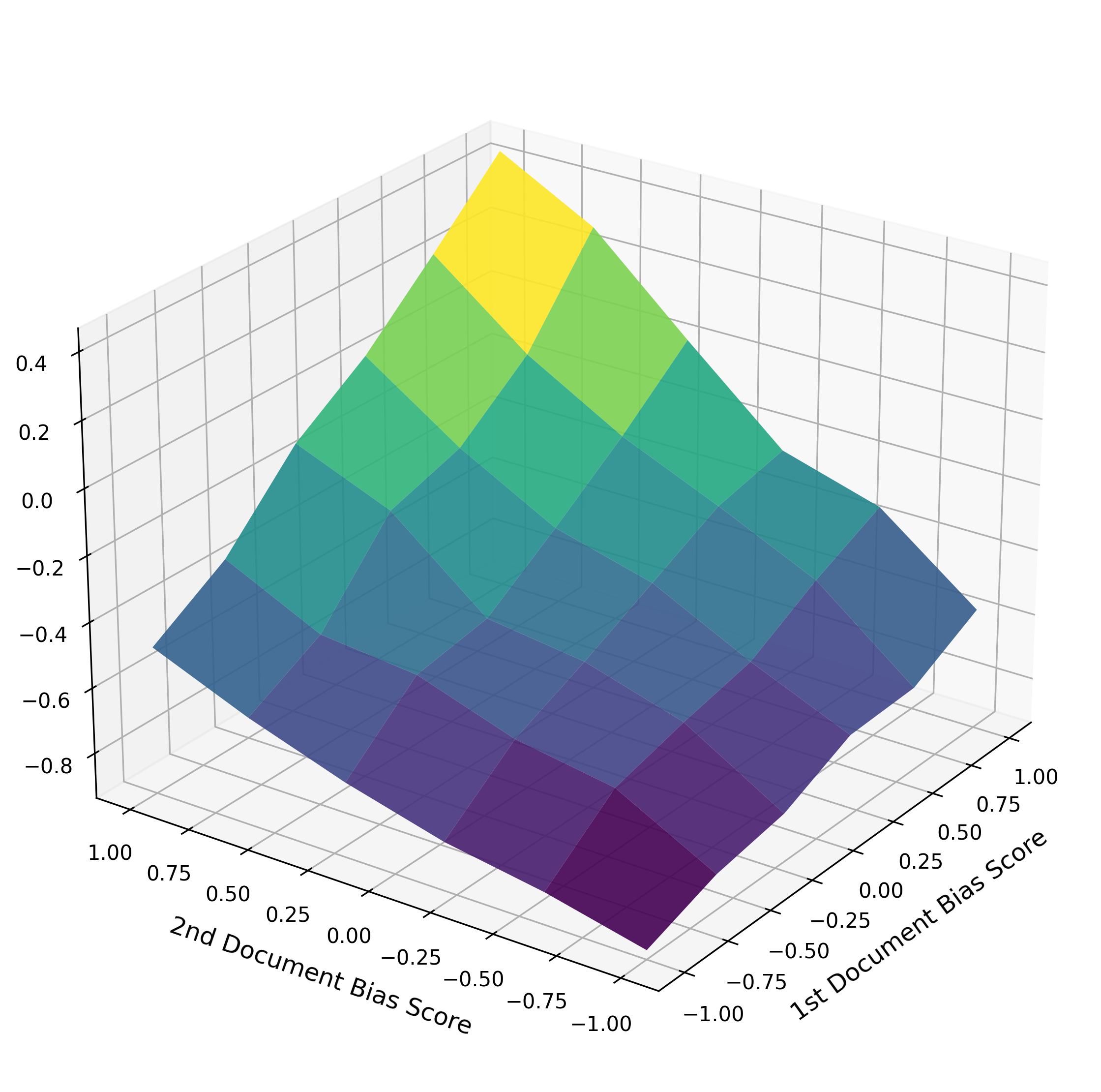}
        \caption{Political: Qwen}
        \label{fig:Qwen}
    \end{subfigure}
    \hfill
    \begin{subfigure}{0.24\linewidth}
        \centering
        \includegraphics[width=\linewidth]{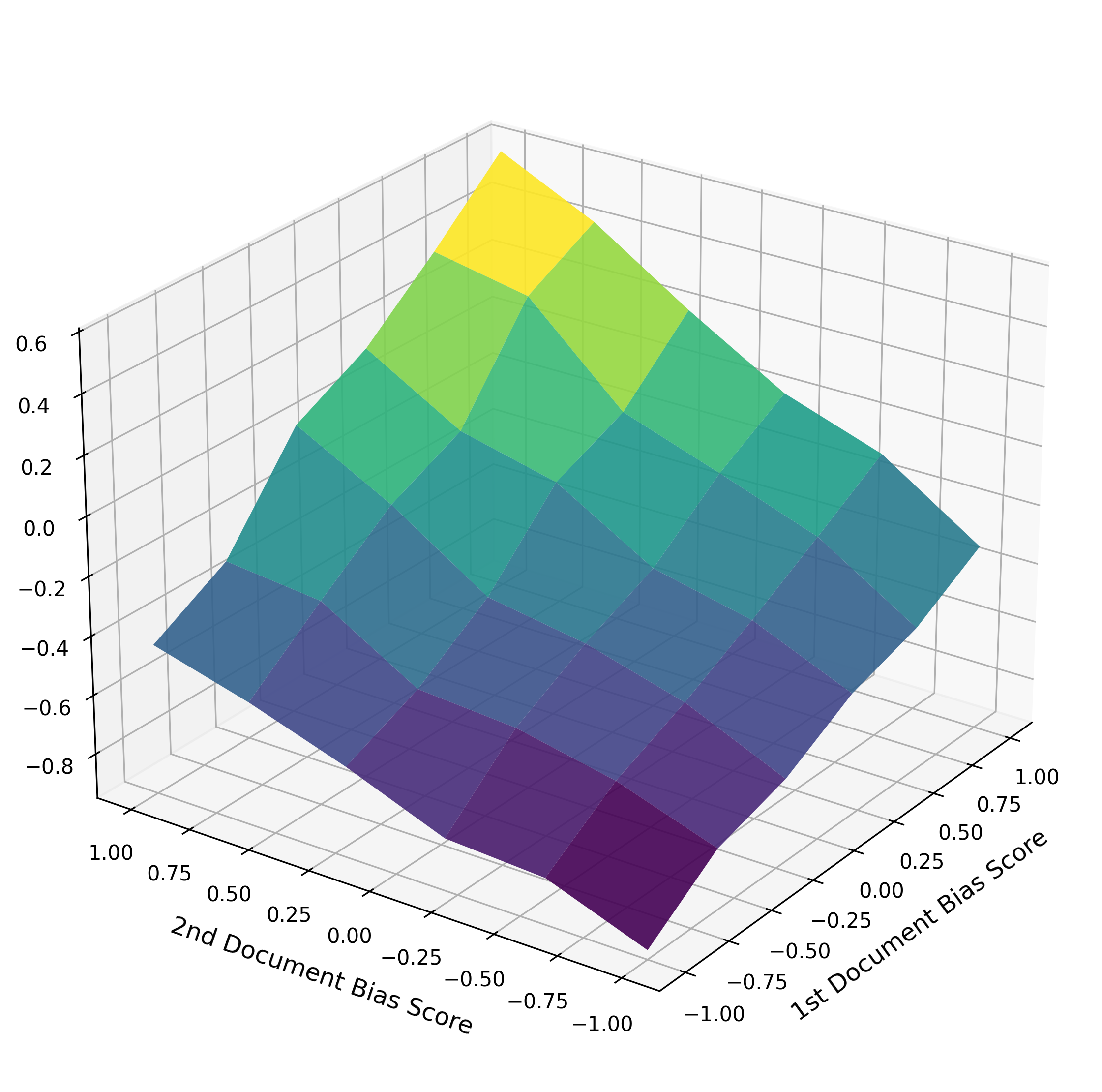}
        \caption{Political: Mistral}
        \label{fig:Mistral}
    \end{subfigure}
    \begin{subfigure}{0.24\linewidth}
        \centering
        \includegraphics[width=\linewidth]{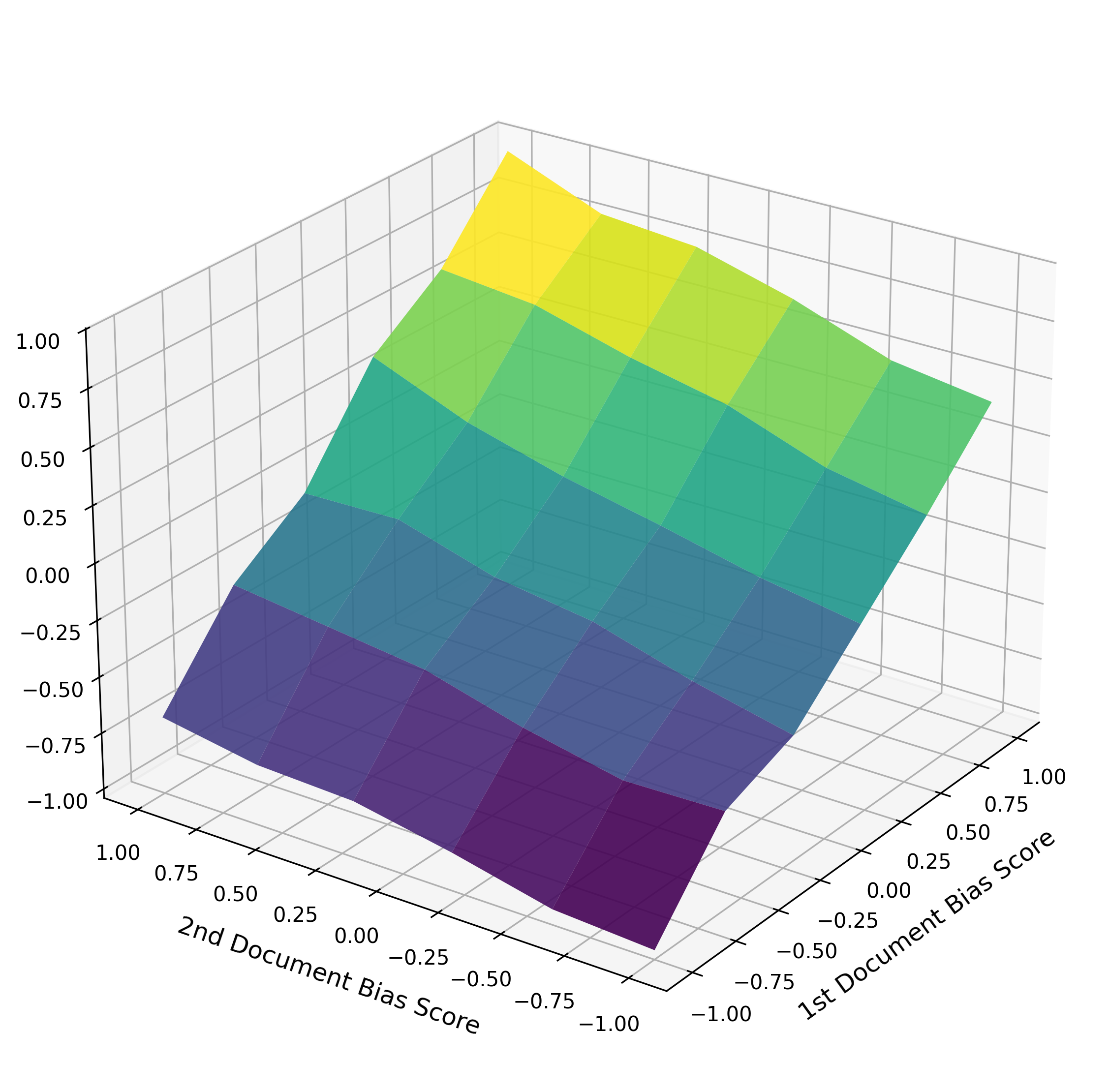}
        \caption{Gender: Llama}
        \label{fig:Llama}
    \end{subfigure}
    \hfill 
    \begin{subfigure}{0.24\linewidth}
        \centering
        \includegraphics[width=\linewidth]{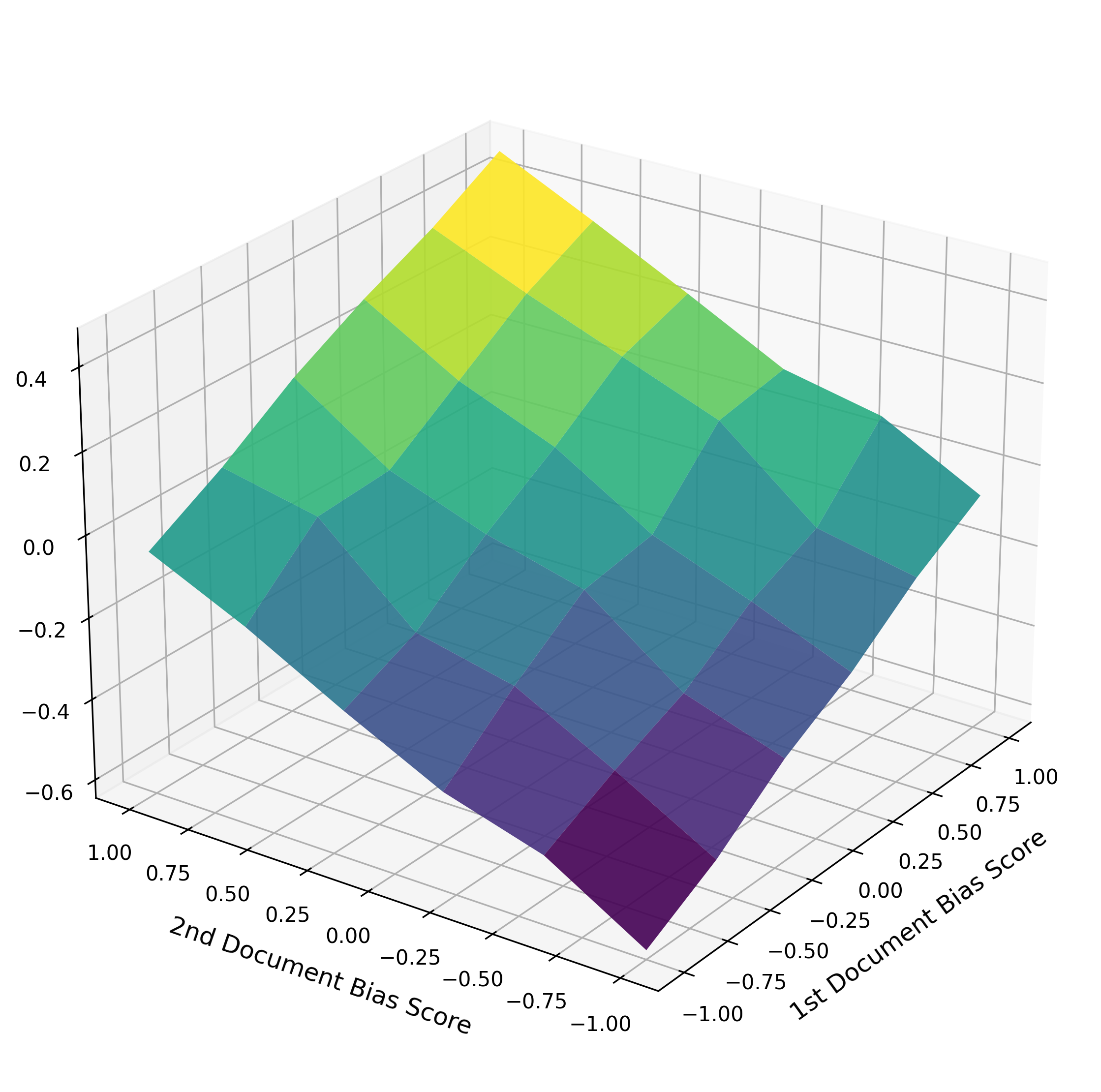}
        \caption{Gender: Gemma}
        \label{fig:Gemma}
    \end{subfigure}
    \hfill
    \begin{subfigure}{0.24\linewidth}
        \centering
        \includegraphics[width=\linewidth]{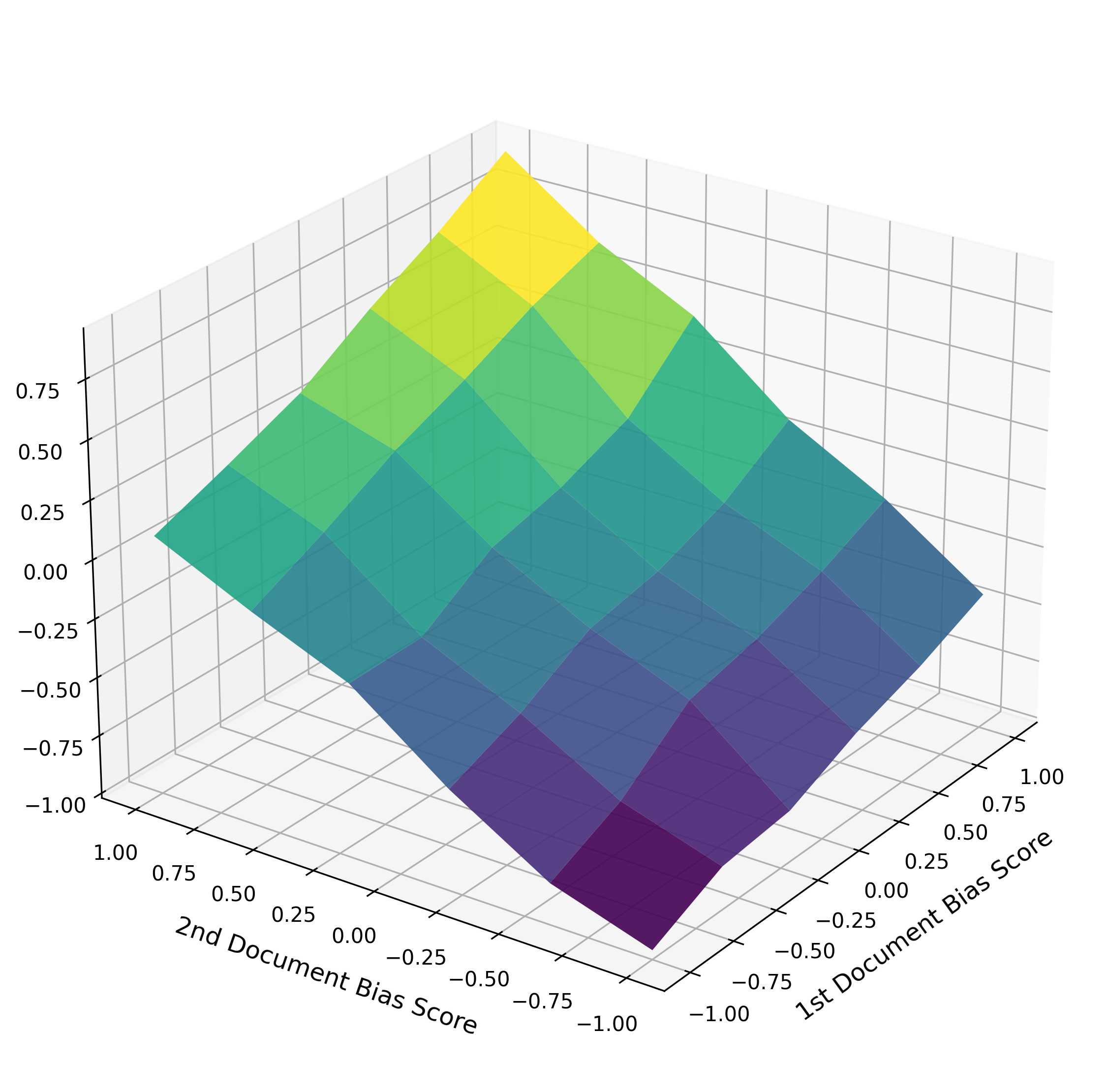}
        \caption{Gender: Qwen}
        \label{fig:Qwen}
    \end{subfigure}
    \hfill
    \begin{subfigure}{0.24\linewidth}
        \centering
        \includegraphics[width=\linewidth]{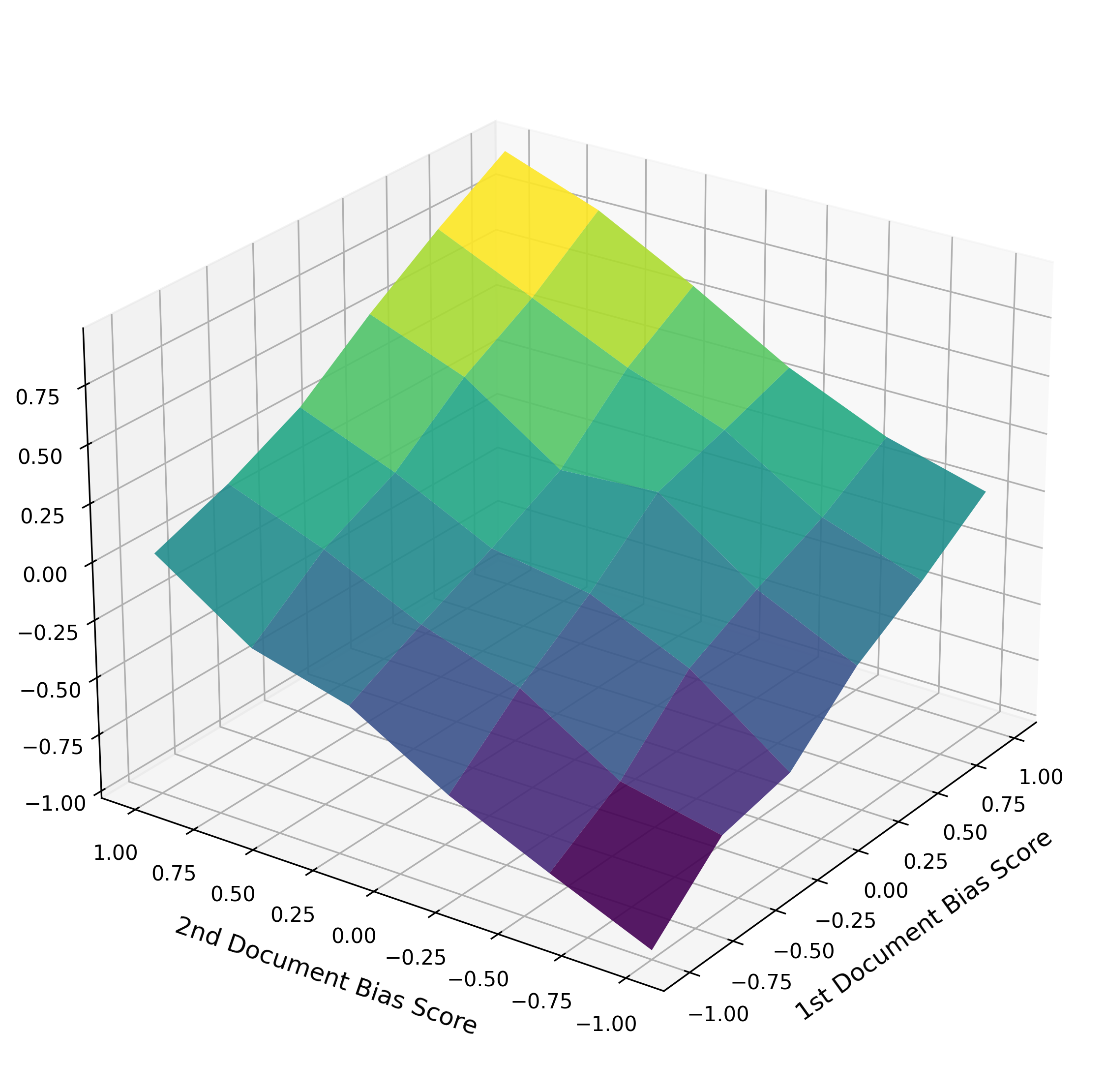}
        \caption{Gender: Mistral}
        \label{fig:Mistral}
    \end{subfigure}
    \caption{Top-2 RAG grid search results: We conducted our grid search experiments based on four popular LLMs. The x and y axes of each graph represent the political and gender embedding bias scores at the first and second positions ($E_b^1$ and $E_b^2$, the value is consistent with the value set in $V_2$), respectively, while the z axis represents the RAG system output bias scores $R_b$ after testing on the corresponding bias dataset.}
    \label{fig:Top-2 RAG grid search results}
\end{figure}

For the top-2 setting, Figure~\ref{fig:Top-2 RAG grid search results} visualizes the relationship between embedding bias at different positions and the resulting output bias for both political and gender biases. Despite differences in how models weigh individual positions, all models exhibit a clear linear trend. 
For higher values of $k$ (top-3 and top-5), direct visualization becomes impractical. 

Therefore, we assess predictive accuracy using five-fold cross-validation
over the bias configurations, separately for each bias type,
LLM, and retrieval depth $k\in\{2,3,5\}$.
Each configuration consists of the position-wise embedding
biases $(E_b^1,\ldots,E_b^k)$ and the observed output bias $R_b$,
aggregated over the corresponding evaluation question set.
We randomly partition the configurations into five folds.
In each iteration, we estimate the position weights and
intercept by ordinary least squares using four folds and
predict the output bias for the remaining fold.
Repeating this procedure yields exactly one out-of-fold
(OOF) prediction for every configuration, obtained from a
model fitted without that configuration.

\begin{figure}[htbp]
    \centering
    \includegraphics[width=\linewidth]
        {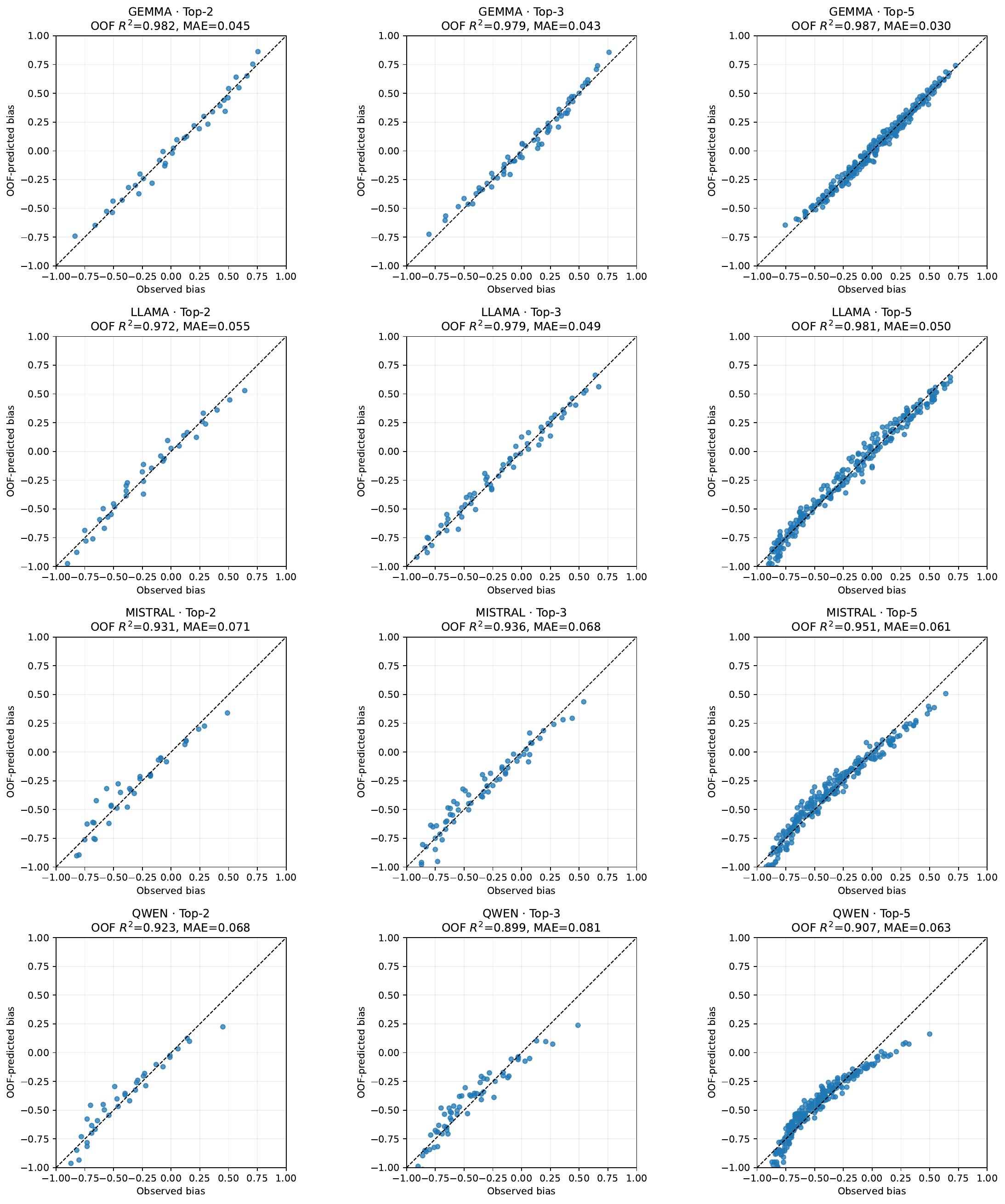}
    \caption{Out-of-fold predicted versus observed RAG output
    bias for political bias across four LLMs and retrieval depths
    $k\in\{2,3,5\}$. Each point represents one retrieval-bias
    configuration, with observed $R_b$ on the horizontal axis
    and its five-fold OOF prediction on the vertical axis.
    The dashed diagonal indicates perfect agreement.
    Reported $R^2$ and MAE are computed over the pooled
    OOF predictions within each setting.}
    \label{fig:poli_predicted_vs_observed}
\end{figure}

\begin{figure}[htbp]
    \centering
    \includegraphics[width=\linewidth]
        {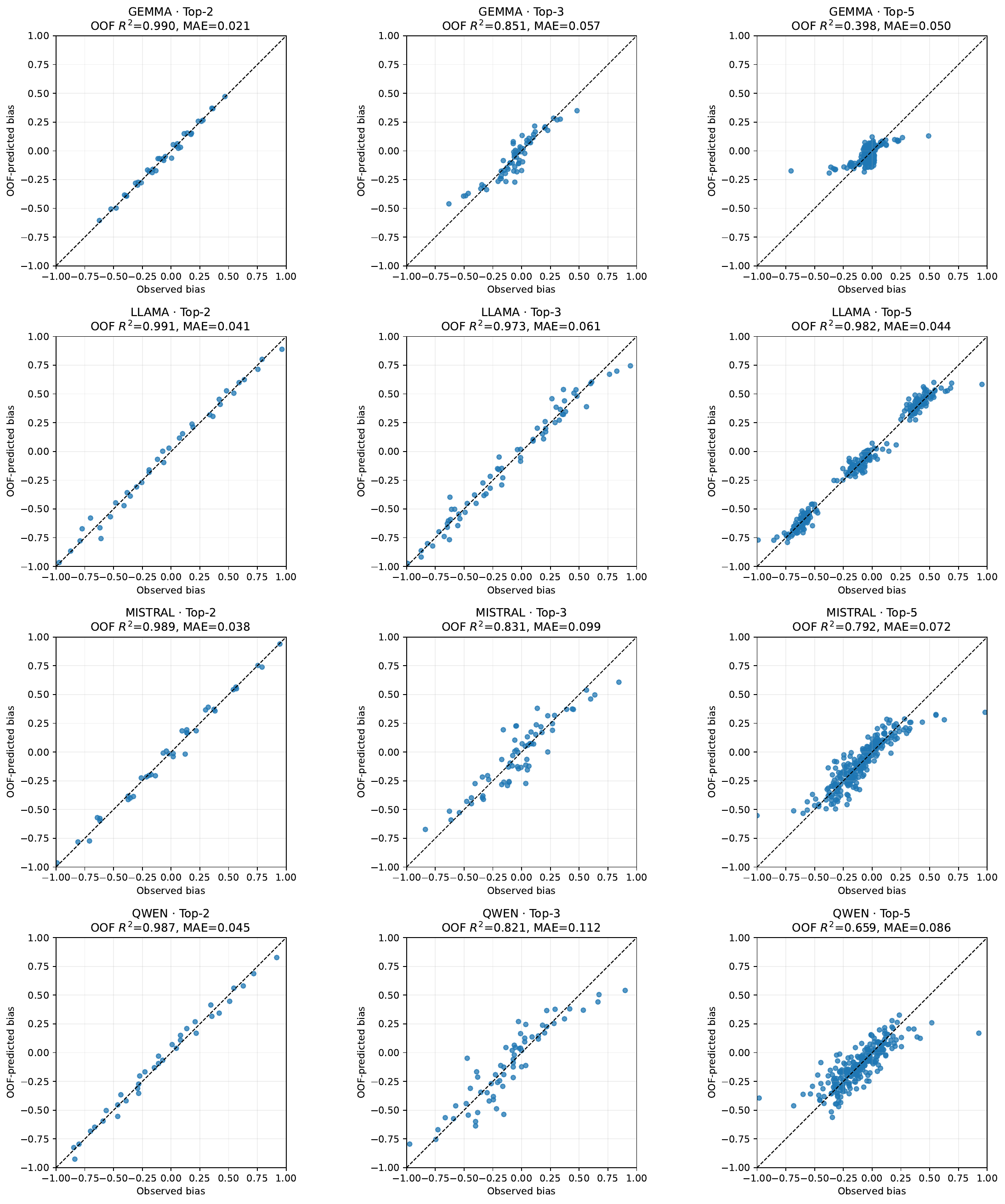}
    \caption{Out-of-fold predicted versus observed RAG output
    bias for gender bias across four LLMs and retrieval depths
    $k\in\{2,3,5\}$. Each point represents one retrieval-bias
    configuration, with observed $R_b$ on the horizontal axis
    and its five-fold OOF prediction on the vertical axis.
    The dashed diagonal indicates perfect agreement.
    Reported $R^2$ and MAE are computed over the pooled
    OOF predictions within each setting.}
    \label{fig:gender_predicted_vs_observed}
\end{figure}

Figure~\ref{fig:poli_predicted_vs_observed} and ~\ref{fig:gender_predicted_vs_observed} compare observed output bias with OOF-predicted bias. Each point represents one configuration, with observed bias on the horizontal axis and predicted bias on the vertical axis. Predictions from all five held-out folds are pooled within each setting, and the dashed diagonal indicates perfect agreement. We compute $R^2$ and mean absolute error (MAE) over the pooled OOF predictions to quantify predictive accuracy. 
{
\begin{align}
    \mathrm{MAE}
    &= \frac{1}{N}\sum_{i=1}^{N}
    \left|R_{b,i}-\widehat{R}_{b,i}^{\mathrm{OOF}}\right|,\\
    R^2
    &= 1-
    \frac{\sum_{i=1}^{N}
    \left(R_{b,i}-\widehat{R}_{b,i}^{\mathrm{OOF}}\right)^2}
    {\sum_{i=1}^{N}
    \left(R_{b,i}-\overline{R}_b\right)^2},
    \qquad
    \overline{R}_b=\frac{1}{N}\sum_{i=1}^{N}R_{b,i}.
\end{align}
} 

Here, $N$ is the number of configurations in the corresponding
bias-type, LLM, and retrieval-depth setting,
$R_{b,i}$ is the observed output bias for configuration $i$,
and $\widehat{R}_{b,i}^{\mathrm{OOF}}$ is its OOF prediction.
MAE measures the average absolute prediction error, while
$R^2$ measures the reduction in squared error relative to
predicting the mean observed bias.
Both metrics are computed over the pooled OOF predictions.
This evaluation tests generalization to unseen bias
configurations within the same evaluation question set.

The political-bias results show close agreement between observed and OOF-predicted output bias, with \(R^2\) ranging from 0.899 to 0.987 and MAE from 0.030 to 0.081. Gemma and Llama exhibit the strongest agreement across all retrieval depths, while Mistral also captures the overall trend with somewhat larger deviations. Qwen shows the largest departures from the diagonal, with predictions tending to underestimate the magnitude of extreme observed biases.

For gender bias, all four LLMs exhibit close agreement between observed and OOF-predicted output bias in the top-2 setting, with \(R^2\) ranging from 0.987 to 0.991. At greater retrieval depths, predictive accuracy becomes more model-dependent. Llama maintains high accuracy across all settings (\(R^2\geq0.973\)); although its top-5 predictions form distinct clusters, these remain largely aligned with the diagonal. Gemma exhibits the largest decline, reaching \(R^2=0.398\) in the top-5 setting. Gemma’s high refusal rate in the gender-bias setting suggests that its responses may be influenced by safeguards concerning gender-related content. Such refusal behavior may partly account for the weaker agreement between observed bias and the linear predictions at greater retrieval depths. Mistral and Qwen retain the overall trend at larger retrieval depths, while showing greater dispersion and underestimation of some extreme bias values. 

These results support the position-aware linear model as a useful approximation for predicting bias under unseen retrieval configurations, while highlighting model-dependent approximation errors.

\begin{figure}[t]
    \centering
    \includegraphics[width=\linewidth]
    {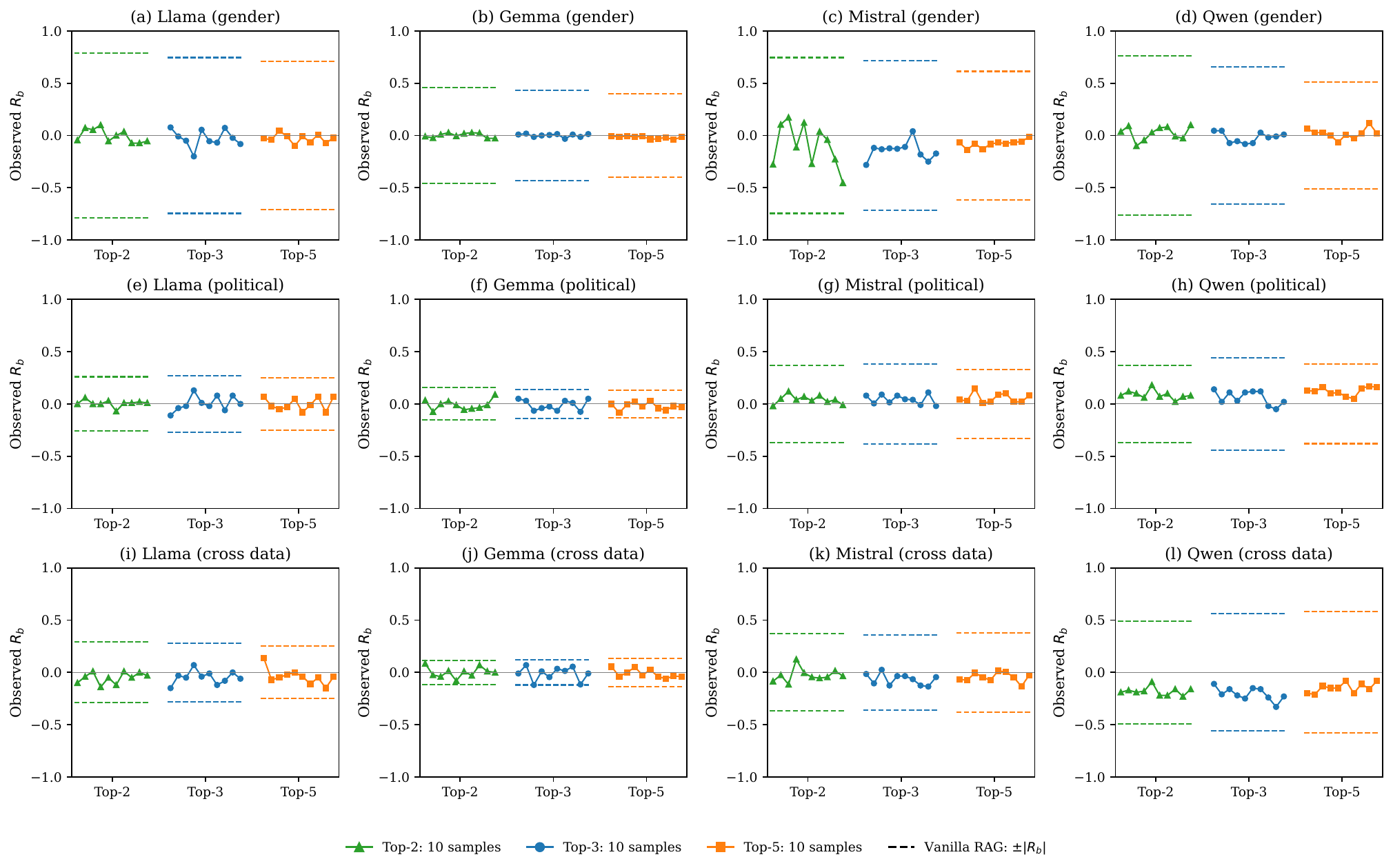}
    \caption{Validation of retrieval configurations sampled
    from the predicted zero-bias hyperplane
    $\widehat{R}_b=0$.
    Rows correspond to gender bias, political bias, and
    cross-dataset political bias; columns correspond to
    Llama, Gemma, Mistral, and Qwen.
    Each panel reports the observed output bias $R_b$
    for 10 sampled configurations at each retrieval depth:
    top-2 (green triangles), top-3 (blue circles), and
    top-5 (orange squares).
    The corresponding dashed lines mark
    $\pm|R_b^{\mathrm{vanilla}}|$ for each retrieval depth.
    Points strictly between these bounds have lower absolute
    output bias than the corresponding vanilla RAG baseline;
    proximity to zero indicates agreement with the
    predicted zero-bias target.}
    \label{fig:verification Rb}
\end{figure}

Additionally, we estimate linear regression models using the dataset constructed in Equation~\eqref{eq:linear input–output pairs} and validate them through sampling. Specifically, we sample configurations of \(E_b^p\) that satisfy the fairness condition \(R_b=0\), and evaluate the resulting bias scores. The results are shown in Figure~\ref{fig:verification Rb}. 
To further assess generalization, we apply the learned linear parameters across different question sets (third row in Figure~\ref{fig:verification Rb}). Despite variations in questions and answer options, the model maintains its effectiveness, with all results remaining within the baseline bias intervals. This indicates that the linear approximation captures stable patterns of bias propagation across datasets.

Overall, these findings provide strong empirical support for the linear bias propagation model and justify its use as the foundation for optimization.

\subsection{Position-wise Bias Analysis}
\label{sec:bias weights results} 
As described in Section 3.2, under top-$k$ settings (k = 2, 3, 5), we perform a grid search over position-wise embedding bias and evaluate the resulting output preference distribution of the RAG system. The goal of this analysis is to understand how different positions in the retrieved context contribute to generation bias, and whether large language models exhibit consistent positional attention patterns across bias types. We then apply linear regression to estimate how much each model relies on documents at different positions, yielding the position-wise sensitivity weights $w_p$ in Equation~\eqref{eq:top-k Rb}.

\begin{figure}[th!]
    \centering
    \begin{subfigure}[b]{0.49\linewidth}
        \centering
        \includegraphics[width=\linewidth]{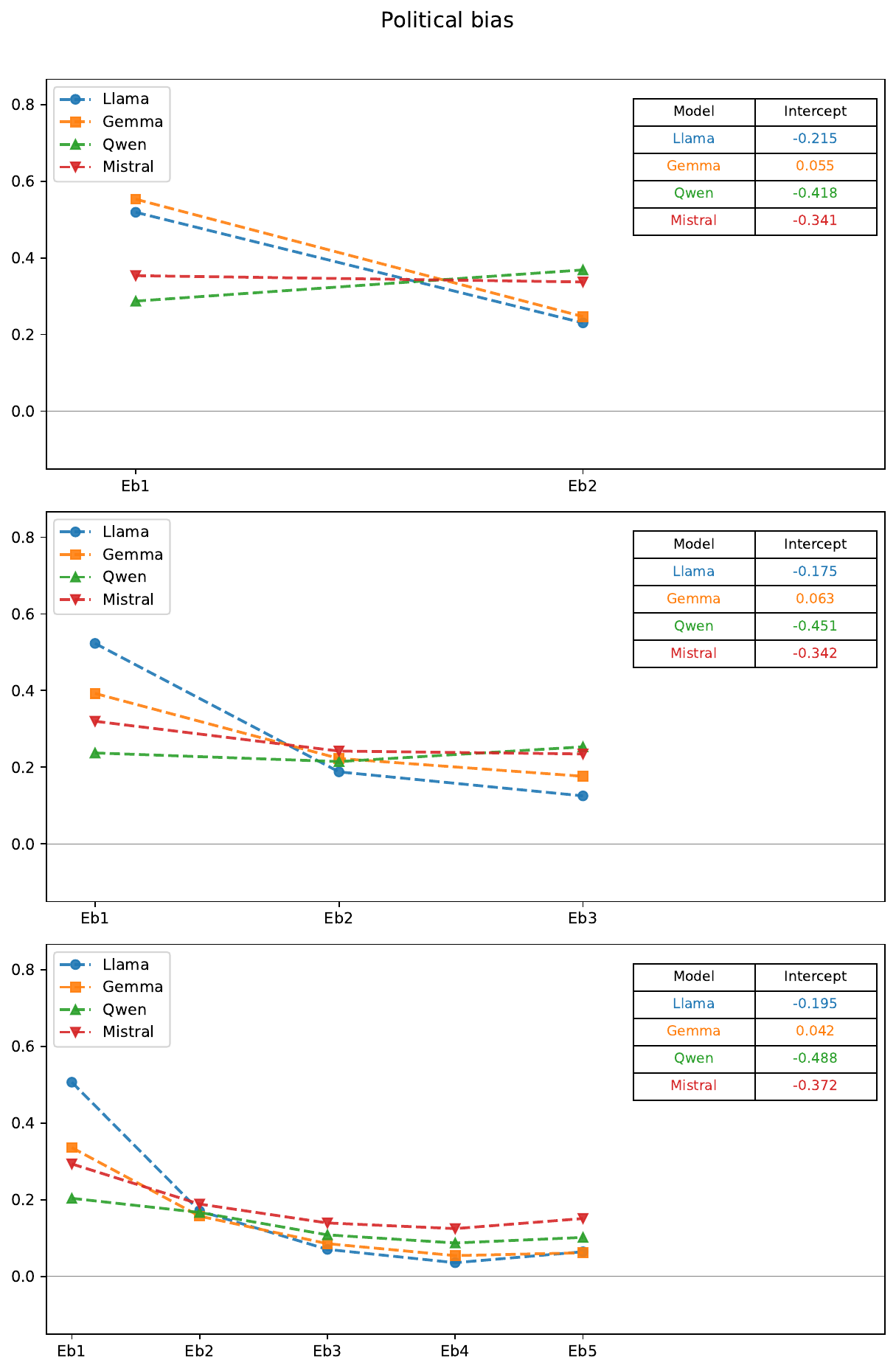}
    \end{subfigure}
    \hfill 
    \begin{subfigure}[b]{0.49\linewidth}
        \centering
        \includegraphics[width=\linewidth]{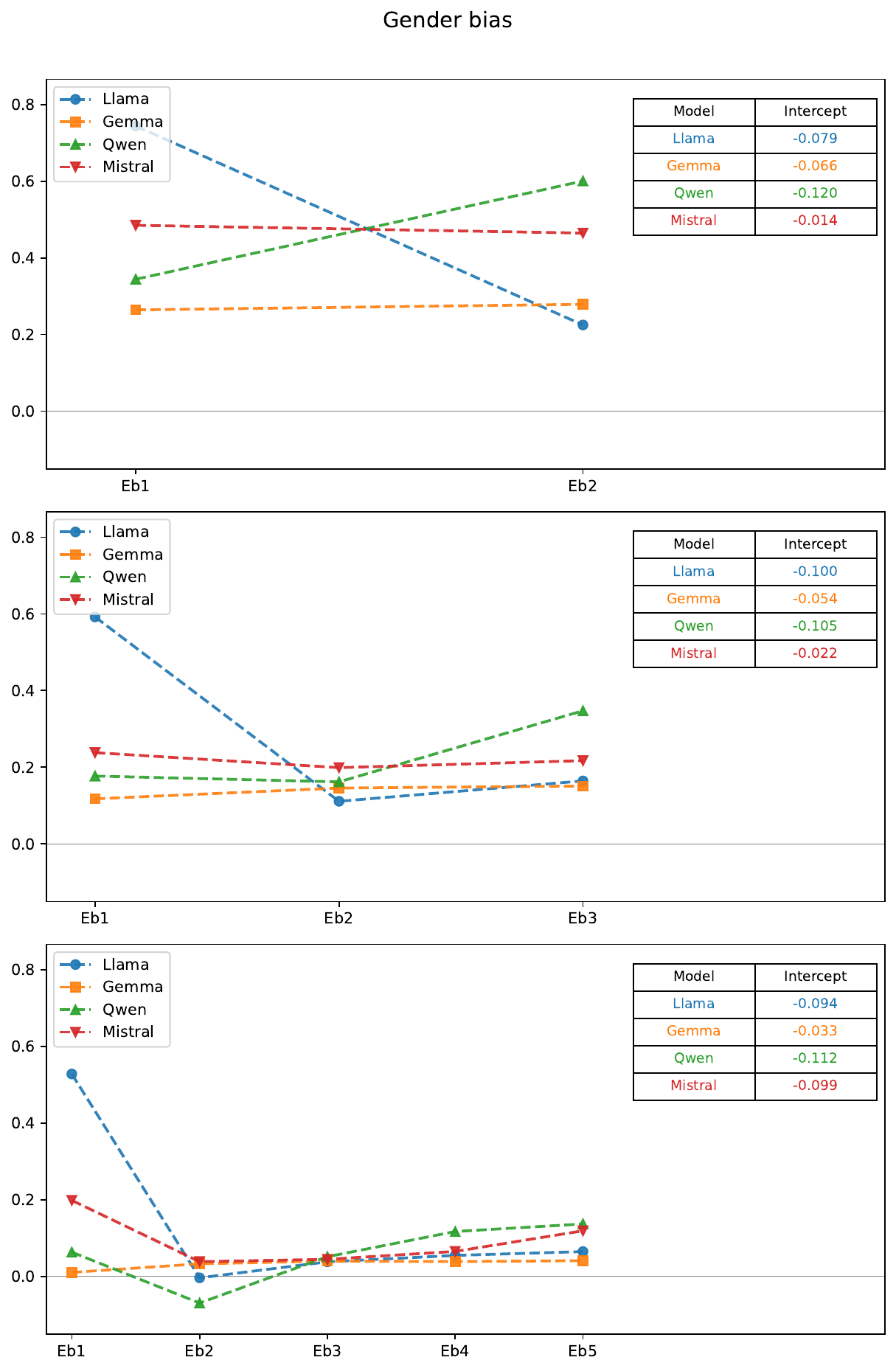}
    \end{subfigure}
    
    \caption{Comparison of position-dependent weight distributions of biased content in the context across LLMs.
Columns correspond to political bias (left) and gender bias (right), while rows correspond to top-$k$ values of $2$, $3$, and $5$ from top to bottom.
Markers denote models: Llama (blue circles, \protect\llamamarker), 
Gemma (orange squares, \protect\gemmamarker), 
Qwen (green upward triangles, \protect\qwenmarker), 
and Mistral (red downward triangles, \protect\mistralmarker).}
    \label{fig:coef_profiles_combined_1}
\end{figure}

To facilitate comparison across models and bias types, we visualize the learned weights in Figure~\ref{fig:coef_profiles_combined_1}. In each plot, the horizontal axis represents the document position, while the vertical axis represents the corresponding weight value. The table within each figure reports the intercept term (i.e., $L_b + \epsilon$) and the corresponding linear model for each LLM.

We observe systematic differences in positional attention across models. For political bias, Llama and Gemma place stronger emphasis on documents at the beginning of the context, indicating a higher reliance on top-ranked evidence. Mistral distributes attention toward both early and late positions, while Qwen allocates attention more evenly across the retrieved documents. 
For gender bias, these patterns shift: Llama continues to prioritize early positions, Qwen shows a stronger preference for later positions, and Gemma and Mistral exhibit more balanced distributions across the context. These results suggest that bias propagation is strongly model-dependent, reinforcing the need to estimate position-wise weights separately for each model.

Across all models, we observe that as $k$ increases, the influence of any single position decreases. This indicates that models rely on a broader aggregation of contextual signals when more documents are included. Consequently, bias control in higher-$k$ settings requires coordinated adjustments across multiple positions, rather than focusing solely on top-ranked documents. 
Interestingly, in the top-5 gender bias experiments, some positional weights become negative. This suggests that documents at certain positions may counteract bias introduced elsewhere in the context, possibly due to redundancy or conflicting information. While the overall bias propagation is well approximated by a linear model, this observation highlights the presence of more complex interactions within the retrieved context. 

The intercept terms further reveal intrinsic model tendencies. In particular, Qwen exhibits a strong preference toward liberal-leaning outputs, consistent with the baseline results reported in Table~\ref{tab:baseline}. Such inherent biases may limit the effectiveness of retrieval-based mitigation, as they introduce a systematic offset that cannot be fully corrected through document selection alone. 

\begin{figure}[th!]
    \centering
    \includegraphics[width=1.0\linewidth]{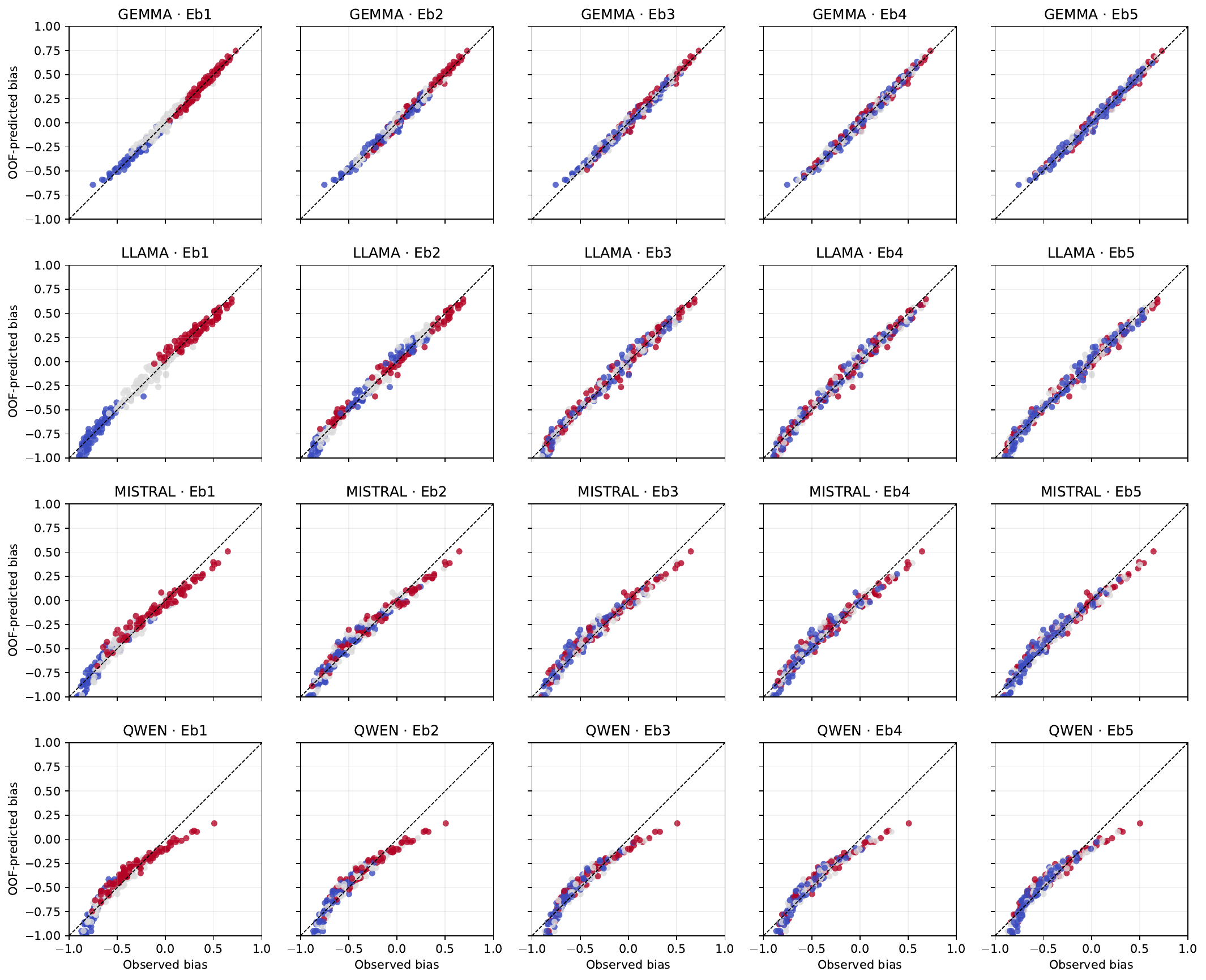}
    \caption{Observed versus out-of-fold predicted political bias for Top-5 RAG configurations, colored by the document bias $E_b^p$ at retrieval position $p$. Rows correspond to models and columns to positions $p=1,\ldots,5$. Each column shows the same configurations with different position-based coloring: blue indicates $E_b^p=-1$, white $0$, and red $1$. The dashed line denotes perfect prediction.}
    \label{fig:poli_top5_position_color}
\end{figure}

\begin{figure}[th!]
    \centering
    \includegraphics[width=1.0\linewidth]{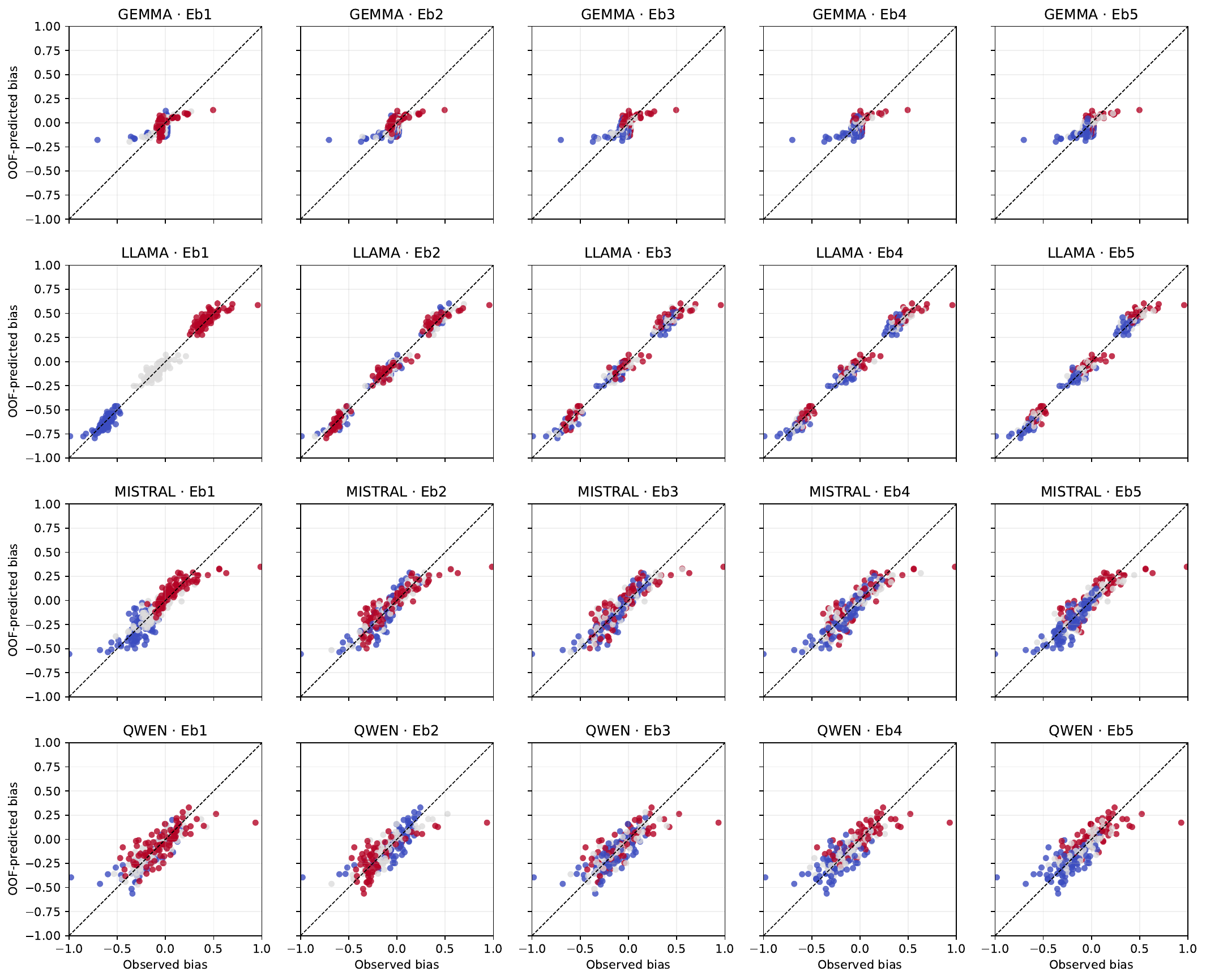}
    \caption{Observed versus out-of-fold predicted gender bias for Top-5 RAG configurations, colored by the document bias $E_b^p$ at retrieval position $p$. Rows correspond to models and columns to positions $p=1,\ldots,5$. Each column shows the same configurations with different position-based coloring: blue indicates $E_b^p=-1$, white $0$, and red $1$. The dashed line denotes perfect prediction.}
    \label{fig:gender_top5_position_color}
\end{figure}

Figures~\ref{fig:poli_top5_position_color} and ~\ref{fig:gender_top5_position_color} visualize the top-5 OOF predictions from Section~\ref{Validation}, coloring each configuration by the position-wise embedding bias \(E_b^p\in\{-1,0,1\}\) at each retrieval position in turn. The resulting color patterns are consistent with the relative magnitudes of the learned position coefficients. For Llama, the larger coefficients at early positions are reflected in distinct clusters that largely follow the values of \(E_b^1\) in both bias settings. Gemma exhibits a similar pattern for political bias, whereas the separation is less pronounced for gender bias. In contrast, Mistral and Qwen distribute their weights more evenly across positions, resulting in less distinct clustering associated with any single position.

\begin{figure}[th!]
    \centering
    \begin{subfigure}[b]{0.49\linewidth}
        \centering
        \includegraphics[width=\linewidth]{v2/political_position_weight_profiles.pdf}
        \label{fig:scaling_gender}
    \end{subfigure}
    \hfill 
    \begin{subfigure}[b]{0.49\linewidth}
        \centering
        \includegraphics[width=\linewidth]{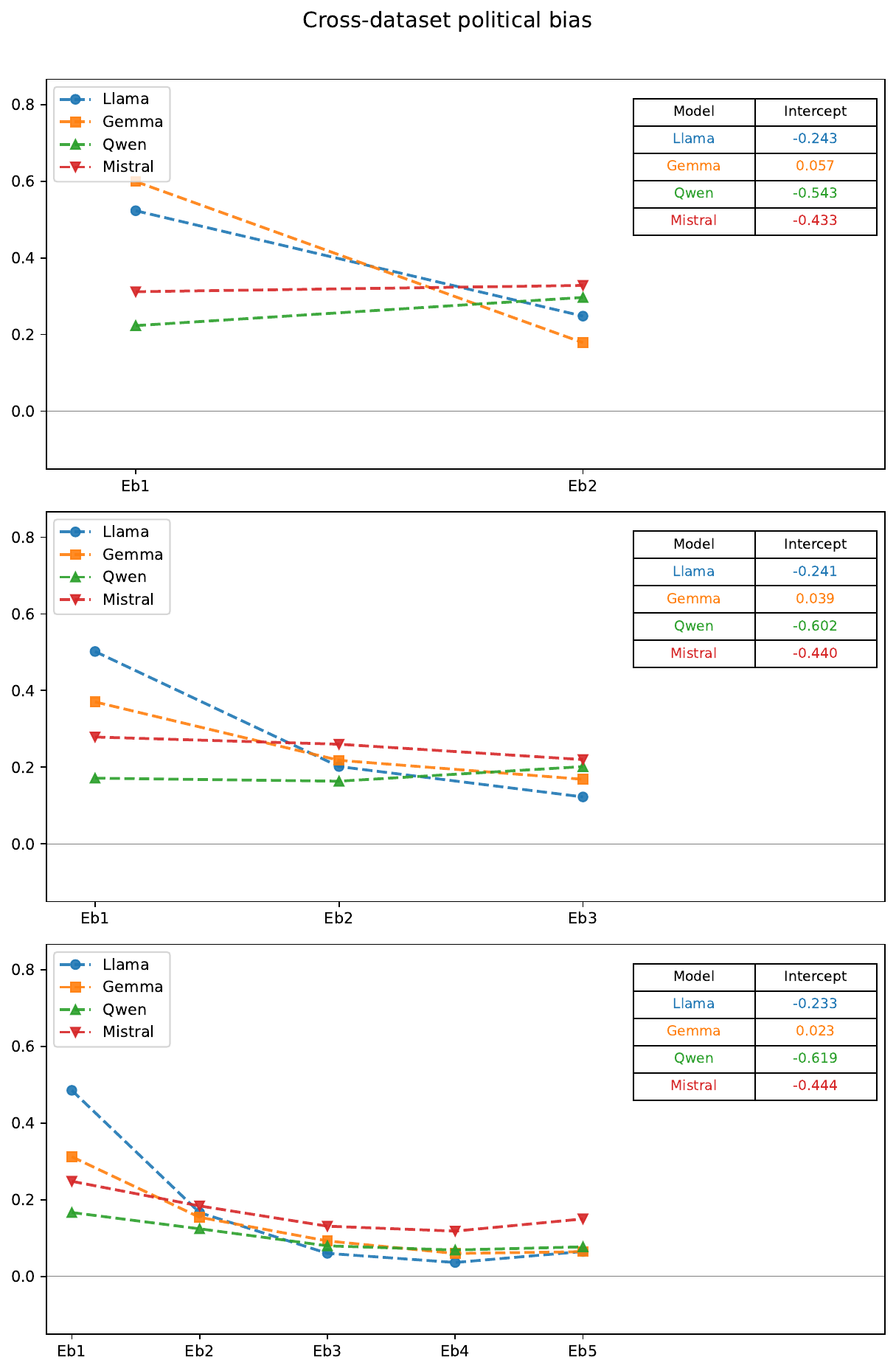}
        \label{fig:scaling_poli}
    \end{subfigure}
    \caption{Comparison of attention weight distributions for political bias across different question sets. 
The layout follows that of the previous figure: rows correspond to top-$k$ values of $2$, $3$, and $5$ (from top to bottom), while columns correspond to different question sets. Markers denote models: Llama (blue circles, \protect\llamamarker), 
Gemma (orange squares, \protect\gemmamarker), 
Qwen (green upward triangles, \protect\qwenmarker), 
and Mistral (red downward triangles, \protect\mistralmarker).}
    \label{fig:coef_profiles_combined_2}
\end{figure}

To assess the robustness of the learned position-wise weights, Figure~\ref{fig:coef_profiles_combined_2} compares the attention distributions obtained from different evaluation question sets. Although the exact numerical values of the weights vary across datasets, the overall positional patterns remain largely consistent. This suggests that the learned bias propagation model captures stable, model-specific behaviors that generalize across datasets. However, such generalization should be applied with caution, as variations in question distributions may still affect mitigation performance.

Overall, these findings confirm that bias propagation in RAG systems is both position-dependent and model-specific, providing strong support for the position-aware optimization framework introduced in Section~\ref{Optimization-Based Fair Retrieval in RAG}.

\subsection{Optimization Results} 
In this section, we evaluate the effectiveness of the proposed FARO algorithm in balancing relevance and fairness, and compare it against the linear programming (LP) baseline. We aim to assess (i) how well the optimization controls bias under different settings, and (ii) the trade-offs between fairness, relevance, and computational efficiency. 

\begin{figure}[t]

    \centering
    \includegraphics[width=1.0\textwidth]{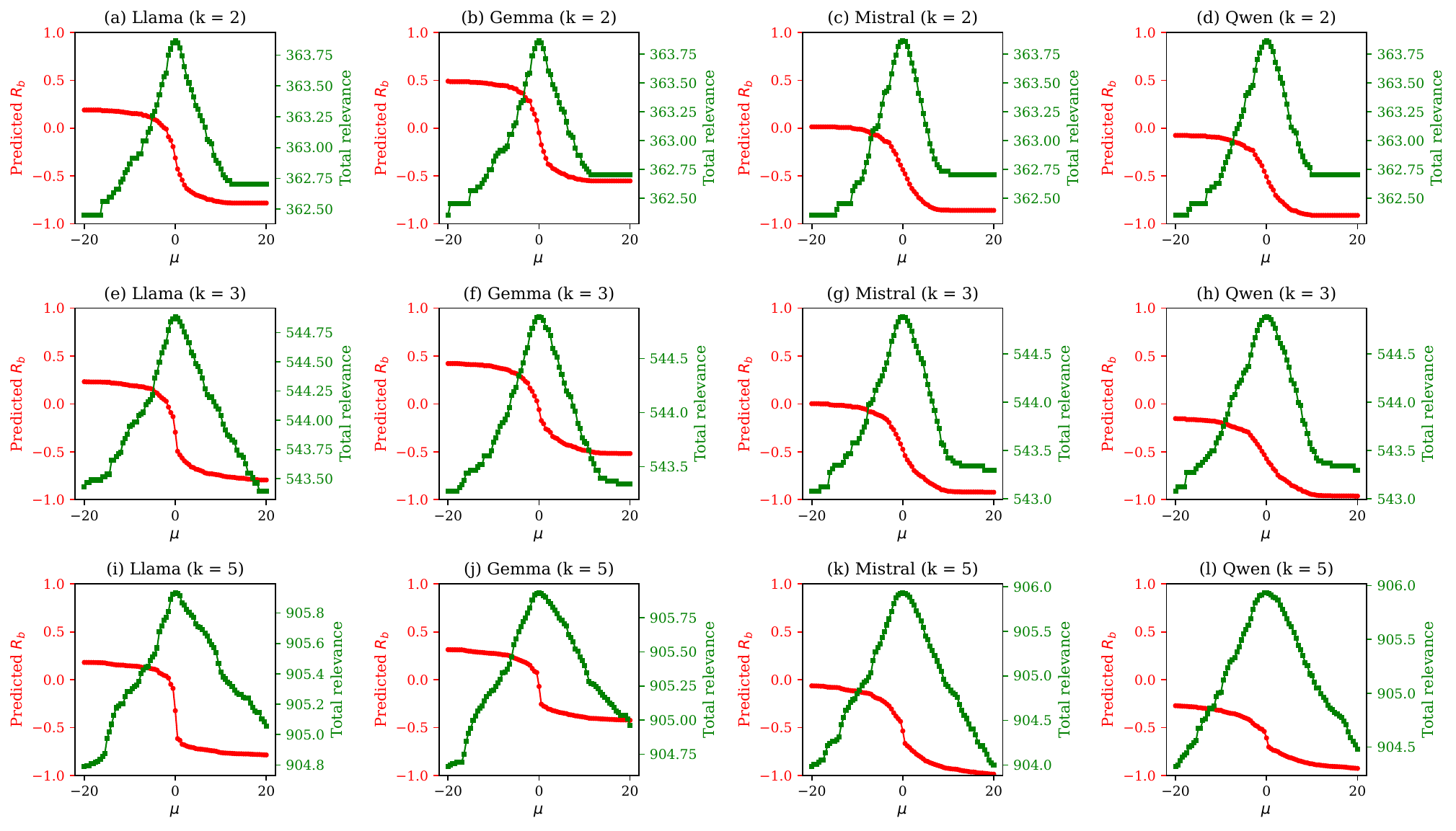}
    \caption{Trade-off curves between fairness and relevance for political bias in RAG systems built on different LLMs. Rows 1 to 3 correspond to top-2, top-3, and top-5 retrieval settings, respectively. The horizontal axis in each plot represents the searched values of $\mu$. The red curve (circular markers, \protect\redcircle) shows the theoretical bias score $R_b$, 
    while the green curve (square markers, \protect\greensquare) represents the actual total relevance score.}
    \label{fig:matrix_trade-off_curve_poli}
\end{figure}

\begin{figure}[t]

    \centering
    \includegraphics[width=1.0\textwidth]{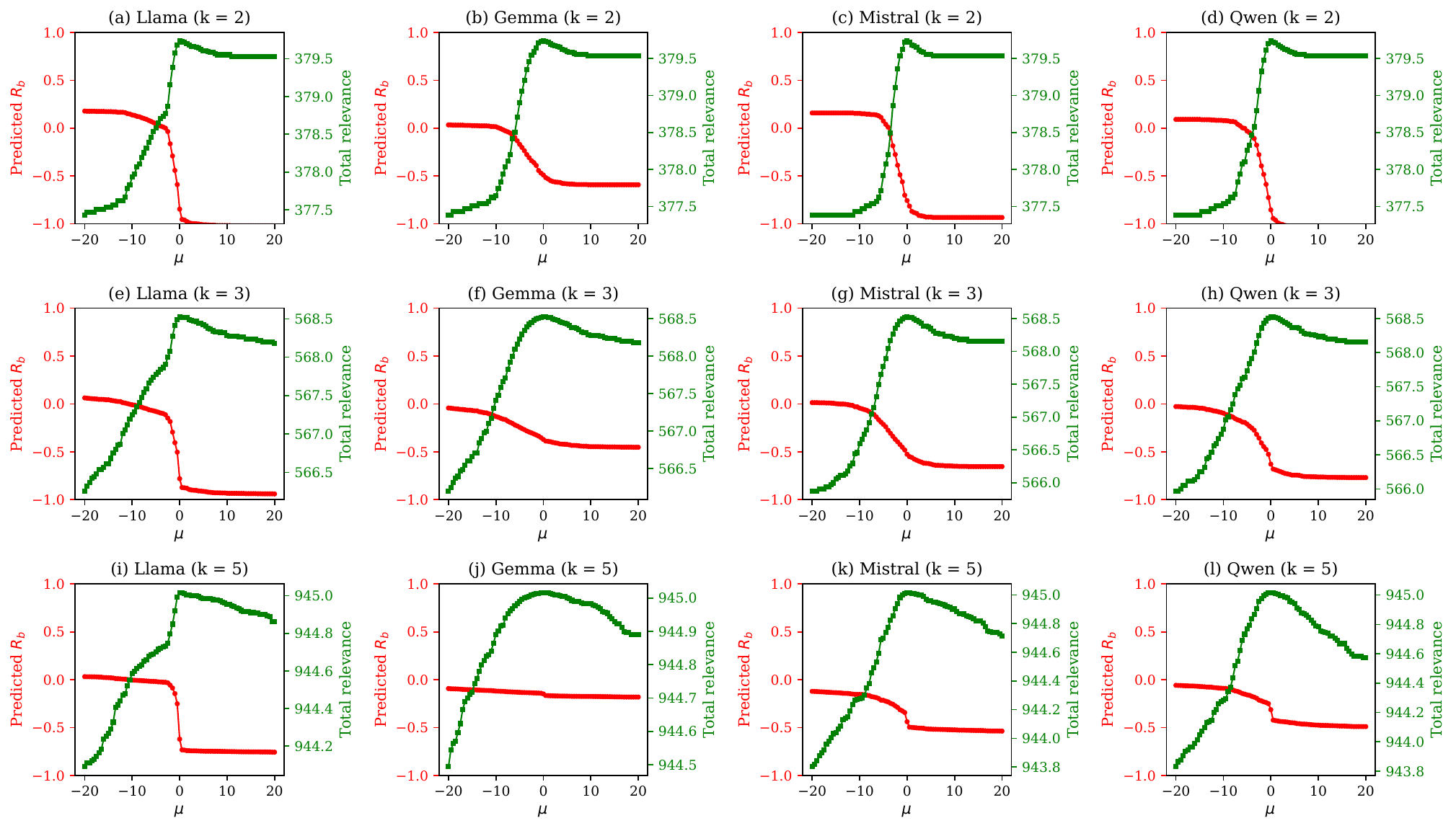}
    \caption{Trade-off curves between fairness and relevance for gender bias in RAG systems built on different LLMs. Rows 1 to 3 correspond to top-2, top-3, and top-5 retrieval settings, respectively. The horizontal axis in each plot represents the searched values of $\mu$. The red curve (circular markers, \protect\redcircle) shows the theoretical bias score $R_b$, 
    while the green curve (square markers, \protect\greensquare) represents the actual total relevance score.}
    \label{fig:matrix_trade-off_curve_gender}
\end{figure}

For the FARO method, we search over 81 uniformly spaced values of $\mu$ in the range [-20, 20], generating a set of candidate retrieval strategies. The results for political and gender bias settings are visualized in Figures~\ref{fig:matrix_trade-off_curve_poli} and~\ref{fig:matrix_trade-off_curve_gender}. In these figures, the horizontal axis represents the values of $\mu$, the red curve (circular markers) shows the theoretical bias score $R_b$, and the green curve (square markers) represents the total relevance score. Together, these curves provide a clear view of the trade-off between fairness and relevance, where different values of $\mu$ correspond to different operating points along the trade-off frontier.

In the political bias setting, the knowledge base provides balanced support for both viewpoints. As a result, the relevance curve exhibits an approximately symmetric (inverted V) shape, peaking near $\mu = 0$. In contrast, the bias curve changes more sharply due to the intrinsic tendencies of the underlying LLMs, which generally favor liberal-leaning outputs. This asymmetry indicates that achieving fairness may require stronger adjustments in retrieval compared to relevance optimization alone.

In the gender bias setting, the knowledge base is skewed toward male-associated content. Consequently, when $\mu < 0$, the optimization favors the underrepresented group, resulting in lower bias but also reduced relevance. Conversely, for $\mu > 0$, the relevance remains relatively high while the bias increases rapidly. This behavior highlights a more pronounced fairness–relevance trade-off compared to the political setting, driven by imbalance in the candidate pool.

Following the predefined fairness constraint $|R_b| \leq 0.1$, we select the retrieval strategy that satisfies the constraint while achieving the highest relevance score. Tables~\ref{tab:political bias opt compare} and~\ref{tab:gender bias opt compare} report the optimization results for political and gender bias, respectively, comparing FARO and LP under different top-$k$ settings. To facilitate comparison, relevance scores are normalized using Min–Max scaling. 
Specifically, for each question, we collect the top 20 most relevant documents as candidates ($|N_q| = 20$) and rank them by relevance. We define the maximum relevance (Max) as the sum of relevance scores of the top-$k$ documents, and the minimum relevance (Min) as the sum for the $bottom-k$ documents. After scaling, the relevance values reported in the table lie in $[0,1]$: values closer to 1 indicate smaller relevance loss, while values closer to 0 indicate larger loss. 
\begin{equation}
    Rel' = \frac{Rel - Min}{Max - Min}
\end{equation}

In addition, the bias scores reported in the table correspond to the actual bias observed after applying the optimized retrieval solution within the RAG system. In contrast, the theoretical bias scores used during optimization all satisfy the fairness constraint. Therefore, these empirical bias values provide a practical reference for evaluating the effectiveness of bias mitigation via the learned linear relationships.

From the optimization results for political bias in Table~\ref{tab:political bias opt compare}, regarding output bias control, the optimization methods satisfy the fairness constraint in most models and settings, including Qwen, despite its substantial intrinsic bias. Gemma is the main exception, showing a weaker linear fit alongside a high refusal rate. Its safety-oriented post-training provides a possible explanation for the refusal behavior, although this connection is not established by our experiments.

In terms of normalized relevance, both methods achieve very similar performance across all settings, with LP showing a slight overall advantage. This difference arises because FARO relies on grid search and thus provides an approximate solution. However, FARO offers greater flexibility by decomposing the global problem into independent per-question subproblems, each of which can be solved optimally. This structure facilitates adaptation to new questions and stricter fairness constraints. For models with substantial intrinsic bias, such as Qwen, as shown by the QWEN-RAG-3 and QWEN-RAG-5 results, LP fails to obtain a solution under the same experimental conditions, whereas FARO yields effective retrieval strategies. 

\begin{table*}[t]
  \centering
  \footnotesize
  \setlength{\tabcolsep}{3pt}
  \caption{Political bias comparison between vanilla RAG, FARO, and LP.}
  \label{tab:political bias opt compare}
  \begin{adjustbox}{max width=\linewidth}
    \begin{tabular}{@{}lc@{\hspace{10pt}}cc ccc ccc@{}}
    \toprule
    \multirow{2}{*}{Model}
    & \multirow{2}{*}{$k$}
    & \multicolumn{2}{c}{Vanilla RAG}
    & \multicolumn{3}{c}{FARO}
    & \multicolumn{3}{c}{LP} \\
    \cmidrule(lr){3-4}
    \cmidrule(lr){5-7}
    \cmidrule(lr){8-10}
    & & Bias & Refusal Rate
    & Bias & Relevance & Time (s)
    & Bias & Relevance & Time (s) \\
    \midrule

    \multirow{3}{*}{Llama}
    & 2 & -0.2600 & 0\%
    & -0.0400 & 0.9837 & 0.1584
    & -0.0500 & 0.9858 & 0.0731 \\
    & 3 & -0.2400 & 0\%
    & -0.0600 & 0.9954 & 0.2022
    & -0.0900 & 0.9963 & 0.1088 \\
    & 5 & -0.2500 & 0\%
    &  0.0400 & 0.9993 & 0.2979
    &  0.0200 & 0.9996 & 0.1980 \\
    \addlinespace

    \multirow{3}{*}{GEMMA}
    & 2 & -0.1450 & 16.5\%
    & -0.1550 & 1.0000 & 0.1551
    & -0.1550 & 1.0000 & 0.0716 \\
    & 3 & -0.1750 & 26.5\%
    & -0.1400 & 1.0000 & 0.2008
    & -0.1900 & 1.0000 & 0.1109 \\
    & 5 & -0.1350 & 33.5\%
    & -0.1300 & 1.0000 & 0.2901
    & -0.1500 & 1.0000 & 0.1949 \\
    \addlinespace

    \multirow{3}{*}{MISTRAL}
    & 2 & -0.3500 & 1\%
    &  0.0500 & 0.9078 & 0.1590
    &  0.0500 & 0.9125 & 0.0703 \\
    & 3 & -0.3600 & 3\%
    &  0.0400 & 0.9095 & 0.3842
    &  0.0600 & 0.9176 & 0.1283 \\
    & 5 & -0.3650 & 4.5\%
    &  0.0800 & 0.8932 & 0.2950
    &  0.0500 & 0.8932 & 0.2129 \\
    \addlinespace

    \multirow{3}{*}{QWEN}
    & 2 & -0.3800 & 0\%
    &  0.1400 & 0.8241 & 0.1596
    &  0.1400 & 0.8284 & 0.0701 \\
    & 3 & -0.4700 & 0\%
    &  0.0500 & 0.8094 & 0.2006
    & \multicolumn{3}{c}{Infeasible} \\
    & 5 & -0.4200 & 0\%
    & -0.0400 & 0.8769 & 0.2916
    & \multicolumn{3}{c}{Infeasible} \\

    \bottomrule
  \end{tabular}

  \end{adjustbox}
    \par\vspace{3pt}
  \begin{minipage}{0.95\textwidth}
    \scriptsize
    \textit{Note:} Vanilla RAG denotes retrieval without fairness-aware
    optimization. Time refers to optimization runtime.
    Infeasible denotes the absence of a feasible LP solution under
    the specified fairness constraint.
  \end{minipage}
\end{table*}

For the gender bias setting (Table~\ref{tab:gender bias opt compare}), the baseline RAG system exhibits substantial bias due to the skewed knowledge base. Under this setting, both FARO and LP achieve significant bias mitigation. However, the relevance loss is more pronounced compared to the political setting, as the candidate pool contains relatively fewer documents representing the underrepresented group. This limits the feasible space of fair solutions and leads to a stronger trade-off between fairness and relevance.

\begin{table*}[t]
  \centering
  \footnotesize
  \setlength{\tabcolsep}{3pt}
  \caption{Gender bias comparison between vanilla RAG, FARO, and LP.}
  \label{tab:gender bias opt compare}
  \begin{adjustbox}{max width=\linewidth}
    \begin{tabular}{@{}lc@{\hspace{10pt}}cc ccc ccc@{}}
    \toprule
    \multirow{2}{*}{Model}
    & \multirow{2}{*}{$k$}
    & \multicolumn{2}{c}{Vanilla RAG}
    & \multicolumn{3}{c}{FARO}
    & \multicolumn{3}{c}{LP} \\
    \cmidrule(lr){3-4}
    \cmidrule(lr){5-7}
    \cmidrule(lr){8-10}
    & & Bias & Refusal Rate
    & Bias & Relevance & Time (s)
    & Bias & Relevance & Time (s) \\
    \midrule

    \multirow{3}{*}{Llama}
    & 2 & -0.8000 & 0\%
    &  0.0182 & 0.8851 & 0.1725
    & -0.0455 & 0.9061 & 0.1639 \\
    & 3 & -0.7550 & 0\%
    & -0.0818 & 0.9320 & 0.2179
    & -0.1636 & 0.9342 & 0.1635 \\
    & 5 & -0.6640 & 0\%
    & -0.1318 & 0.9879 & 0.3140
    & -0.1455 & 0.9893 & 0.2530 \\
    \addlinespace

    \multirow{3}{*}{GEMMA}
    & 2 & -0.4860 & 43.2\%
    & -0.0864 & 0.8357 & 0.1678
    & -0.0864 & 0.8377 & 0.1022 \\
    & 3 & -0.4680 & 48.6\%
    & -0.0955 & 0.8475 & 0.4174
    & -0.1273 & 0.8560 & 0.1854 \\
    & 5 & -0.4180 & 53.6\%
    & -0.2500 & 0.9715 & 0.3242
    & -0.2773 & 0.9737 & 0.3418 \\
    \addlinespace

    \multirow{3}{*}{MISTRAL}
    & 2 & -0.7640 & 2.7\%
    &  0.0773 & 0.8337 & 0.1766
    &  0.0227 & 0.8474 & 0.0994 \\
    & 3 & -0.6230 & 3.2\%
    & -0.0409 & 0.8523 & 0.2221
    & -0.0227 & 0.8552 & 0.1729 \\
    & 5 & -0.4950 & 5\%
    & -0.1545 & 0.9118 & 0.3320
    & -0.1636 & 0.8725 & 0.3384 \\
    \addlinespace

    \multirow{3}{*}{QWEN}
    & 2 & -0.7550 & 0\%
    & -0.0182 & 0.8300 & 0.1712
    & -0.0455 & 0.8407 & 0.0975 \\
    & 3 & -0.6450 & 0\%
    & -0.0818 & 0.8362 & 0.2203
    & -0.1273 & 0.8405 & 0.1659 \\
    & 5 & -0.4820 & 0\%
    & -0.0909 & 0.9531 & 0.4862
    & -0.0727 & 0.9541 & 0.3033 \\

    \bottomrule
  \end{tabular}
    \end{adjustbox}
  \par\vspace{3pt}
  \begin{minipage}{0.95\textwidth}
    \scriptsize
    \textit{Note:} Vanilla RAG denotes retrieval without fairness-aware
    optimization. Time refers to optimization runtime.
  \end{minipage}
\end{table*}

\subsection{Runtime and Scalability Analysis}

\begin{figure}[th!]
    \centering
    \captionsetup[subfigure]{
        font=small,
        justification=centering,
        skip=4pt
    }

    \begin{subfigure}[t]{\linewidth}
        \centering
        \includegraphics[width=\linewidth]
            {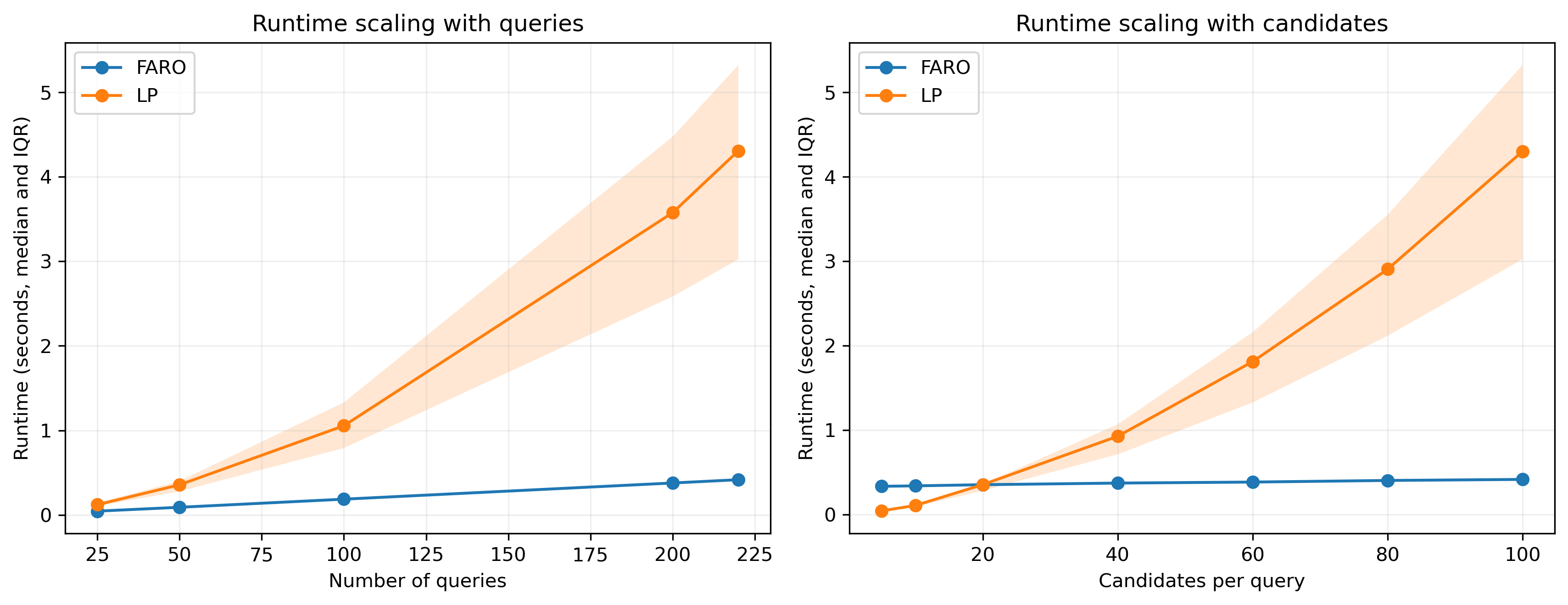}
        \caption{Gender bias}
        \label{fig:scaling_gender}
    \end{subfigure}

    \par\medskip

    \begin{subfigure}[t]{\linewidth}
        \centering
        \includegraphics[width=\linewidth]
            {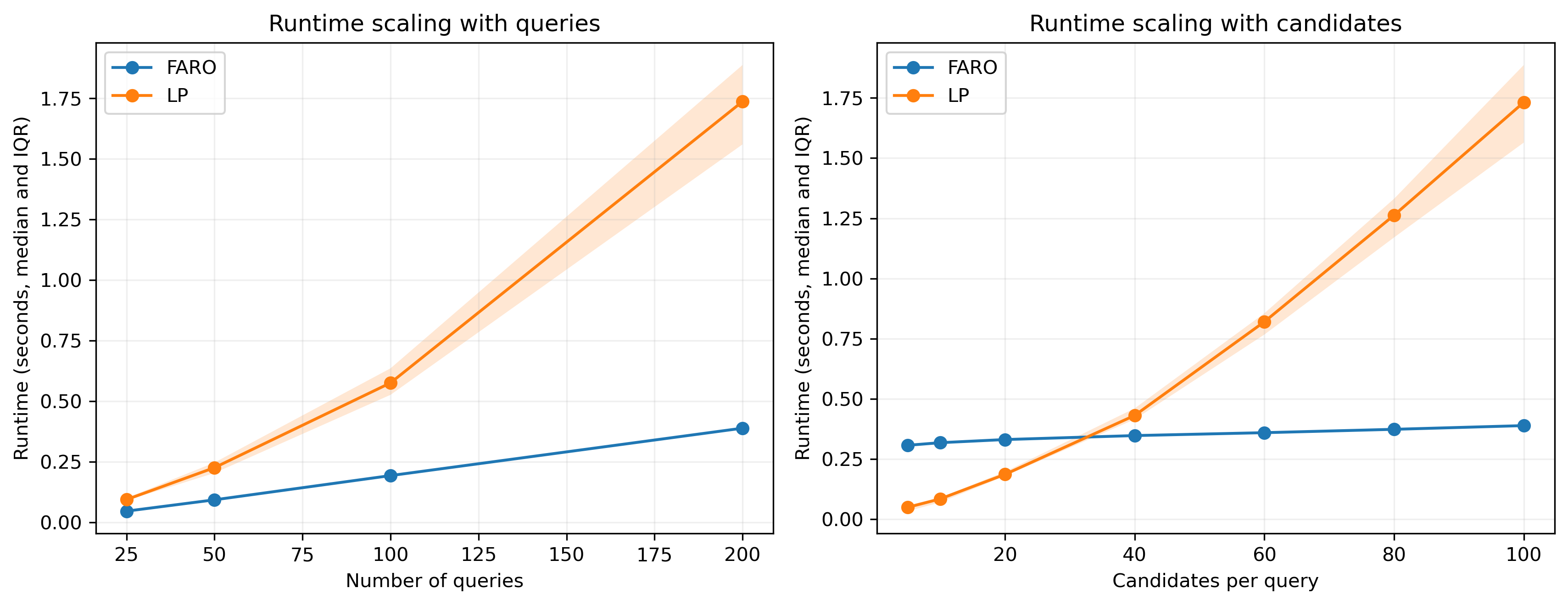}
        \caption{Political bias}
        \label{fig:scaling_poli}
    \end{subfigure}

    \caption{Runtime scaling of FARO and LP under gender and
    political bias. Within each panel, the left plot varies
    the number of queries $|S|$, while the right plot varies
    the candidate pool size $|N_q|$.
    Results aggregate runs using the top-5 linear
    bias-propagation models of all four LLMs, with 30 repetitions
    per setting. Curves show median runtimes, and shaded regions
    indicate the interquartile range (25th--75th percentiles).}
    \label{fig:faro_vs_lp_scaling}
\end{figure}

Given the relatively small candidate pools and question sets used in our main experiments, LP exhibits a clear runtime advantage. To evaluate the scalability of FARO, we measure runtime while independently varying the number of queries and the candidate pool size. For query scaling, we use \(|S|\in\{25,50,100,200\}\), with 100 candidates per query. For candidate scaling, we use \(|N_q|\in\{5,10,20,40,60,80,100\}\), retaining the full evaluation question set: 220 queries for gender bias and 200 for political bias. For each bias type, we evaluate both methods using the top-5 linear bias-propagation models estimated for all four LLMs, with 30 repetitions per setting. Figure~\ref{fig:faro_vs_lp_scaling} reports median runtimes, with shaded regions indicating the interquartile range (25th–75th percentiles).

With 100 candidate documents per query, LP is slightly slower than FARO even with only 25 queries in both bias settings. As the number of queries increases, LP runtime rises sharply, whereas FARO exhibits more gradual growth. When varying the candidate pool size using the full question set, LP is faster for pools smaller than 40 documents, demonstrating its advantage on small problems. Beyond 40 candidates, however, its runtime increases substantially, while FARO maintains comparatively moderate growth. The wider interquartile bands for LP indicate greater runtime variability across the evaluated propagation models and repeated runs, whereas FARO exhibits more consistent runtime.

\subsection{Discussion} 
Overall, the proposed framework achieves effective bias control in RAG systems while maintaining high relevance. 
From an optimization perspective, LP performs well for small-scale problems with fixed constraints, providing exact solutions. However, its scalability limitations make it less suitable for larger or dynamic settings. In contrast, FARO offers a flexible and efficient alternative by decomposing the global problem into independent subproblems and generating multiple candidate solutions along the fairness–relevance frontier. 
This flexibility is particularly valuable in practical applications, where fairness requirements may vary across scenarios. By adjusting the search range and granularity of $\mu$, FARO allows practitioners to balance computational efficiency and solution quality.

Overall, these results demonstrate that combining a position-aware bias model with a scalable optimization framework enables effective and practical fairness-aware retrieval in RAG systems.

\section{Conclusions} 
\label{sec:conclusion}
In this paper, we addressed the problem of fairness in top-k RAG, focusing on how retrieval decisions influence bias in generated outputs. We proposed a unified three-stage framework that enables controlled bias injection, models position-aware bias propagation, and optimizes retrieval under fairness constraints. 

Our approach introduced a linear, position-aware bias propagation model that captures how multiple retrieved documents jointly affect system-level bias. Building on this model, we formulated fairness-aware retrieval as an optimization problem and proposed the FARO algorithm, which enables flexible exploration of the relevance–fairness trade-off. Experimental results across multiple datasets and models demonstrate that: (i) bias in RAG systems can be effectively controlled through retrieval, (ii) the proposed linear model provides a strong approximation of bias propagation in top-k settings, and (iii) the FARO algorithm achieves competitive performance compared to linear programming while offering improved scalability and flexibility.  

Despite these promising results, several limitations remain. First, the proposed bias propagation model relies on linearity and independence assumptions, which may not fully capture complex interactions between retrieved documents. Second, our formulation focuses on binary group fairness and does not directly extend to multi-group fairness definitions. Third, the effectiveness of bias mitigation depends on the intrinsic biases of the underlying language models. 
Future work includes extending the framework to richer fairness notions, incorporating non-linear or interaction-aware bias models, and exploring adaptive retrieval strategies that account for question-specific fairness requirements.

\section*{Declaration of generative AI and AI-assisted technologies
in the manuscript preparation process}

During manuscript preparation, the authors used AI tools
to assist with Chinese-to-English translation. The authors reviewed and revised all
AI-assisted material, verified its accuracy, and take full
responsibility for the final manuscript.

\section*{Acknowledgment}
This work has received funding from the European Union’s Horizon Europe research and innovation programme through the GANNDALF project under grant agreement No.101167951. Views and opinions expressed are those of the author(s) only and do not necessarily reflect those of  the European Union.

The authors wish to acknowledge CSC - IT Center for Science,
Finland, for computational resources.

\bibliographystyle{elsarticle-num-names} 
\bibliography{biblio}

\appendix




\end{document}